\documentclass[aip,superscriptaddress,twocolumn,10pt]{revtex4}%
\usepackage{amsmath}
\usepackage{amsfonts}
\usepackage{amssymb}
\usepackage{graphics}
\usepackage{graphicx}
\providecommand{\U}[1]{\protect\rule{.1in}{.1in}}
\begin{document}
\title{Weakly Flux-Tunable Superconducting Qubit}

\author{Jos\'{e} M. Ch\'{a}vez-Garcia}
\thanks{J.M.C-G, F.S., B.A. contributed equally to this work. \\ J.M.C-G current address: Center for Quantum Devices, Niels Bohr Institute, University of Copenhagen, 2100 Copenhagen, Denmark.}
\author{Firat Solgun}
\author{Jared B. Hertzberg}
\author{Oblesh Jinka}
\author{Markus Brink}
\author{Baleegh Abdo}
\email{babdo@us.ibm.com}
\affiliation{IBM Quantum, IBM Research Center, Yorktown Heights, New York 10598, USA.}
\date{\today}

\begin{abstract}	
Flux-tunable qubits are a useful resource for superconducting quantum processors. They can be used to perform cPhase gates, facilitate fast reset protocols, avoid qubit-frequency collisions in large processors, and enable certain fast readout schemes. However, flux-tunable qubits suffer from a trade-off between their tunability range and sensitivity to flux noise. Optimizing this trade-off is particularly important for enabling fast, high-fidelity, all-microwave cross-resonance gates in large, high-coherence processors. This is mainly because cross-resonance gates set stringent conditions on the frequency landscape of neighboring qubits, which are difficult to satisfy with non-tunable transmons due to their relatively large fabrication imprecision. To solve this problem, we realize a coherent, flux-tunable, transmon-like qubit, which exhibits a frequency tunability range as small as $43$ MHz, and whose frequency, anharmonicity and tunability range are set by a few experimentally achievable design parameters. Such a weakly tunable qubit is useful for avoiding frequency collisions in a large lattice while limiting its susceptibility to flux noise. 
\end{abstract}

\maketitle

\newpage

\section{Introduction}
\begin{figure}
	[tb]
	\begin{center}
		\includegraphics[
		width=\columnwidth 
		]%
		{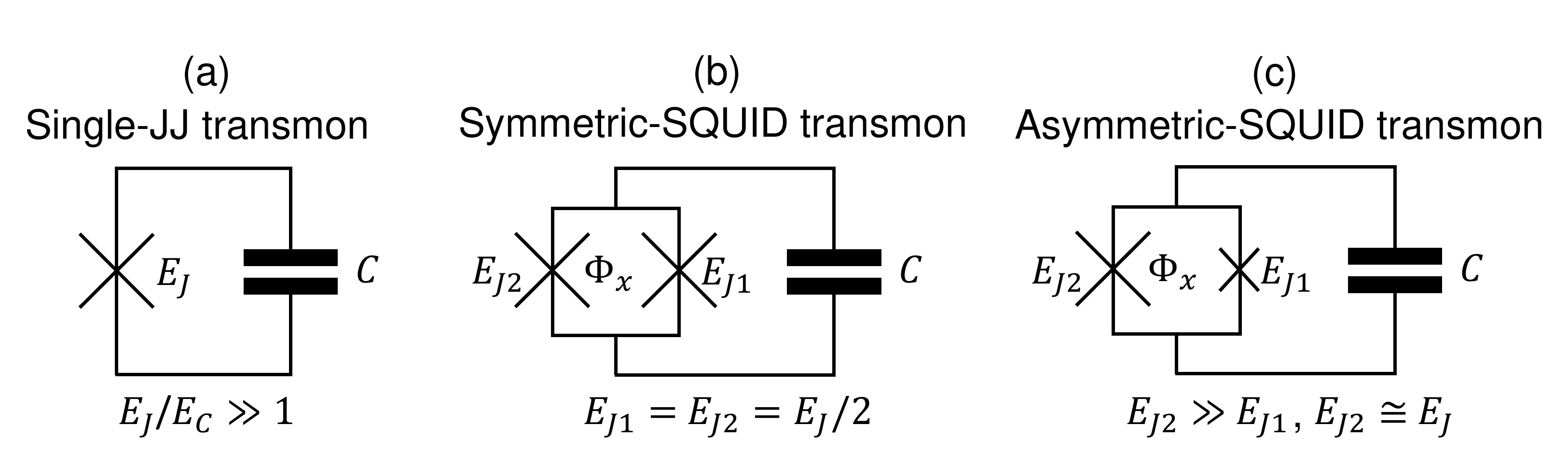}
		\caption{Transmon circuits. (a) Fixed-frequency transmon consisting of a single Josephson junction with energy $E_J$ shunted by a capacitor $C$. (b) Widely tunable transmon consisting of a symmetric dc-SQUID with identical junctions $E_{J1}=E_{J2}=E_{J}/2$ shunted by a capacitor $C$. The qubit frequency can be tuned using an external magnetic flux $\Phi_{x}$ threading the SQUID loop. (c) Medium-range tunable transmon consisting of an asymmetric dc-SQUID with dissimilar junctions $E_{J2} \gg E_{J1}$, where $E_{J2} \cong E_{J}$, shunted by a capacitor $C$.   
		}
		\label{Transmons}
	\end{center}
\end{figure}

Quantum computers promise significant speedup, over their classical counterparts, for certain hard computational problems, such as factoring and quantum chemistry \cite{factoring,ValidatingQC,QChemistry}. However, for quantum computers to achieve a clear advantage over classical computers they need to run error correction codes and have sufficient quantum volumes \cite{QuantumVolume}. One leading architecture for realizing such universal quantum computers is a crystal-like lattice of Josephson-junction-based qubits that supports the surface code or similar variations \cite{SCreview,PlanarCodes}. But to realize such a generic architecture, it is critical to employ high-coherence qubits that are simple to fabricate and characterize, and high-fidelity two-qubit gates that are fast, easy to tune up, and, preferably, require a minimal hardware overhead. Two leading candidates that have been shown to satisfy these requirements are single Josephson-junction (JJ) transmons \cite{transmon,3Dtransmon,3DtransRigetti} and cross-resonance gates, which are fully controlled by microwave signals \cite{CRthy,CRexp}. In particular, single-JJ tansmons, formed by capacitively shunting a JJ (see Fig.\,\ref{Transmons}(a)), exhibit coherence times on the order of a few hundreds of microseconds \cite{ibmpeekskill}, and cross-resonance gates, which realize cNOT gates by generating a ZX-like interaction between two coupled qubits, regularly yield fidelities in excess of $99.1\%$ with gate times of about $300$ ns \cite{CRfidelity}.

However, despite these favorable properties and numerous successful realizations of small quantum processors consisting of tens of single-JJ transmons and cross-resonance gates, deploying such qubits and gates in large quantum processors can be quite challenging. This is because cross-resonance gates, which are based on driving the control qubit at the target qubit frequency, set stringent lower and upper bounds on the first and second energy-level detunings of not only the control and target qubits but also their direct neighbors \cite{laserAnnealing,laserAnnealing2,MultiqubitCorcoles,MultiqubitMaika,OptFreqAlloc}. Satisfying these lower and upper bounds, which are necessary to avoid frequency collisions and slow gates, respectively, is particularly difficult to accomplish with single JJ-transmons. The difficulty arises from the fact that their fixed frequencies $f_q$ are primarily determined by the JJ energies, which owing to uncontrolled parameters in the fabrication process, have random scatter with a standard deviation $\sigma_f$ that is comparable to the upper bounds of the required detunings (set by the qubit anharmonicity). Such imprecision in the occurring transmon frequencies significantly increases the likelihood of frequency collisions between neighboring qubits and decreases the yield of collision-free chips. For example, a `heavy hexagon' type lattice of qubits in a three-frequency pattern was shown to most effectively evade frequency-collisions \cite{laserAnnealing}. Yet even the smallest-sized such lattice, containing 23 qubits, if fabricated with conventional precision of $\sigma_f / f_q \sim 3\%$, will be collision-free only $0.1$\% of the time \cite{laserAnnealing}. 

To address this crippling frequency-collision problem in medium and large quantum processors, several strategies are being pursued, including: 1) replacing the single JJ-transmons with symmetric/asymmetric dc-SQUID transmons (see Fig.\,\ref{Transmons}(b),(c)), whose frequency is tunable with external flux \cite{transmon,AsymmonsExp}; 2) combining single-JJ transmons with large-anharmonicity qubits, such as capacitively-shunted flux qubits \cite{CSfluxqubit}; and 3) selectively modifying undesired transmon frequencies following fabrication and testing by etching a portion of their capacitive pads \cite{APSabstract} or illuminating their JJs with focused laser beams for short durations \cite{laserAnnealing,laserAnnealing2}. The first strategy enables precise qubit frequency tuning and the second eases frequency crowding, but in either case the qubits will suffer dephasing when tuned away from their flux-insensitive `sweet spots.' The third strategy requires complex processing that is challenging to apply to large 3D-integrated and packaged processors. The best demonstrated precision of such schemes is $\sigma_f = 14$ MHz, which can yield a collision-free 23-qubit device $70$\% of the time, but enables only $8$\% yield of a $127$-qubit device and $\ll 0.1\%$ yield at $1000$ qubit scale \cite{laserAnnealing}.

Here, we realize a superconducting qubit, named weakly tunable qubit (WTQ), that retains the desirable properties of single-JJ transmons in multi-qubit architectures and whose frequency tunes weakly with applied magnetic flux. Such a limited tunability could solve the frequency-collision problem in multi-qubit architectures while maintaining high coherence. This tunability could also improve qubit relaxation times by evading two-level systems (TLSs) in frequency space. According to Ref. \cite{DynamicsT1}, shifting a qubit by about $10$ MHz can decouple it from the TLS and restore the qubit’s $T_1$. Moreover, such a qubit could be beneficial in realizing high-fidelity parametric gates that rely on frequency modulation \cite{PGwithTunableTrans,PGanalysis}.   

Prior to introducing the WTQ circuit, we briefly highlight the drawbacks of existing tunable transmons, namely, the symmetric dc-SQUID transmon and the asymmetric dc-SQUID transmon, whose circuits are shown in Fig.\,\ref{Transmons} (b),(c). Since the frequency tunability range in the symmetric case can be large exceeding a gigahertz, its flux-noise sensitivity, which, to first order, increases with $|\rm{d}\textit{f}_q/\rm{d}\Phi_{\textit{x}}|$, results in a significant dephasing away from the sweet spot \cite{AsymmonsExp}. In the asymmetric case on the other hand, a smaller tunability and sensitivity to flux noise can be achieved with large JJ-area ratio. But even with a ratio as high as $15$ to $1$, the smallest achievable tunability is about $330$ MHz, which is much larger than what is needed to avoid frequency collisions. Furthermore, since the Josephson energy of the large-size JJ in the asymmetric SQUID (i.e., $E_{J2}$) is comparable to that of the single-JJ transmon (i.e., $E_{J}$), a significantly thicker oxide is required in the fabrication process to yield $E_{J2}\gg E_{J1}$ (see Fig.\,\ref{Transmons}(c)). Such a thick oxide requirement increases the probability of lossy two-level systems in the JJs, potentially limiting the qubit lifetime. 

The outline of the remainder of the paper is as follows. In Sec. II, we introduce the WTQ circuit. In Sec. III, we derive the WTQ Hamiltonian. In Sec. IV, we present analytical formulas for the frequency and anharmonicity of the WTQ and calculated response of a WTQ example. In Sec. V. we calculate the relaxation and dephasing rates of WTQs. In Sec. VI, we present experimental results, i.e., spectroscopy and coherence, taken of two 7-qubit chips that incorporate 6 WTQs and a fixed-frequency transmon each. In Sec. VII, we discuss the measurement results, offer additional theoretical predictions, and outline possible enhancements and future directions. Finally, in Sec. VIII, we provide a brief summary and highlight the advantages of employing WTQs in large quantum processors.  

\section{The WTQ circuit}

\begin{figure}
	[tb]
	\begin{center}
		\includegraphics[
		width=\columnwidth 
		]%
		{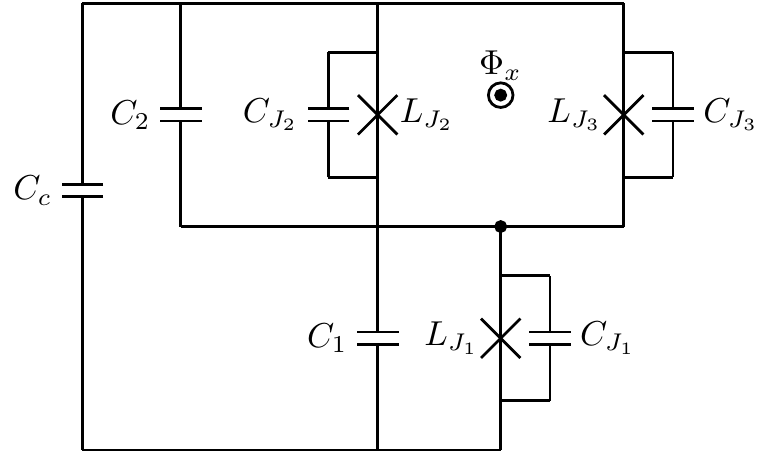}
		\caption{Circuit diagram of the WTQ. It consists of three Josephson junctions with inductances
			$L_{J_{1}}$, $L_{J_{2}}$, $L_{J_{3}}$ and self-capacitances $C_{J_{1}}$,
			$C_{J_{2}}$, $C_{J_{3}}$, respectively. The critical currents of
			the junctions are related to their inductances by the relation $I_{c,i}=\frac{{\Phi_{0}}}{2\pi L_{J_{i}}}$,
			for $i=1,2,3$. The SQUID loop formed by the junctions $J_{2}$ and
			$J_{3}$ is biased by the external DC flux $\Phi_{x}$. The sensitivity
			of the SQUID loop to the flux noise is reduced by making the areas
			of the junctions $J_{2}$ and $J_{3}$ unequal, i.e. $I_{c,3}=\alpha_J I_{c,2}$
			for some $\alpha_J>1$. Capacitors $C_{1}$ and $C_{2}$ shunting the
			junctions create the two main modes of the circuit: the qubit mode
			formed by $J_{1}$ shunted with $C_{1}$ and the high-frequency mode
			formed by $C_{2}$ shunting the SQUID loop. The electrostatic interaction
			of the qubit mode with the SQUID through the capacitance $C_{c}$
			provides the tunability of the WTQ.      
		}
		\label{fig:WTQ-0}
	\end{center}
\end{figure}

The WTQ circuit consists of three Josephson
junctions $J_{1}$, $J_{2}$, $J_{3}$ with self-capacitances $C_{J_{1}}$,
$C_{J_{2}}$, $C_{J_{3}}$, respectively, and three capacitors $C_{1}$,
$C_{2}$ and $C_{c}$ as shown in Fig.\,(\ref{fig:WTQ-0}). The junctions
$J_{2}$ and $J_{3}$ form a SQUID loop which is connected in series
with the junction $J_{1}$. The junction $J_{1}$ shunted by the capacitance
$C_{1}$ provides the main transmon mode of the qubit. The SQUID shunted
by the capacitance $C_{2}$ generates a second transmon-type mode
whose frequency is tuned by the external flux bias $\Phi_{x}$ threading
the SQUID loop. Asymmetry is introduced in the SQUID by making the
areas of the junctions $J_{2}$ and $J_{3}$ unequal to reduce the
sensitivity to the flux noise \cite{AsymmonsExp}. The tunability of the
qubit mode is achieved by the electrostatic interaction of the junction
$J_{1}$ with the SQUID through the capacitance $C_{c}$.

In general, the underlying physics of WTQs is very similar to that of tunable coupling qubits (TCQs) \cite{
	tcqThy,tcqExp,tcqExp2,tcqSuppShotNoise}, which also consist of two capacitively coupled transmons. TCQs and WTQs, however, differ in their circuit and functionality. While TCQs employ nominally identical symmetric SQUID transmons, WTQs employ a single-JJ and asymmetric SQUID transmons. TCQs also allow to independently tune their frequency and coupling strength to the readout resonator, whereas WTQs mainly enable their frequency to be tuned within a small range.
 
\section{Derivation of the WTQ Hamiltonian}

\begin{figure}
	[tb]
	\begin{center}
		\includegraphics[
		width=\columnwidth 
		]%
		{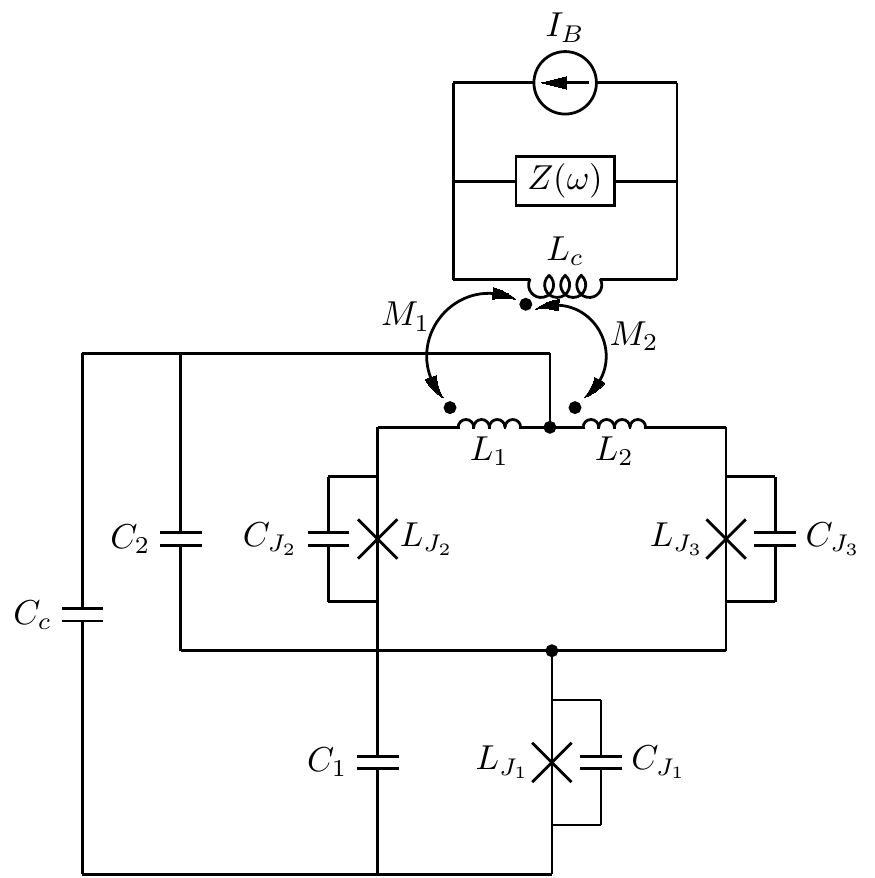}
		\caption{Detailed WTQ Circuit Diagram. The flux bias in
			the SQUID loop is generated by the DC current source $I_{B}$ with
			impedance $Z(\omega)$. $L_{c}$ is the inductance of the coil that
			produces the magnetic field threading the SQUID loop. Two partial
			inductances \cite{Ruehli} $L_{1}$ and $L_{2}$ are introduced to
			model the linear inductance of the SQUID loop which are inductively
			coupled to the coil with mutual inductances $M_{1}$ and $M_{2}$,
			respectively. Such a finely detailed construction of the WTQ circuit
			avoids inconsistenties in the calculation of the decoherence rates  
		}
		\label{fig:WTQ-IB}
	\end{center}
\end{figure}

We will apply the circuit quantization formalism developed in \cite{BKD}
to derive the Hamiltonian of the WTQ circuit and will employ the analysis
technique described in \cite{Brito} to make a Born-Oppenheimer approximation
and calculate the decoherence rates. To better describe the coupling of the
WTQ to the flux bias circuitry we introduce a more detailed
circuit model of the flux bias mechanism as shown in Fig.\,(\ref{fig:WTQ-IB}).
The flux bias in the SQUID loop is generated by the DC current source
$I_{B}$ with impedance $Z(\omega)$. Note that this construction
is quite general in the sense that a wide range of circuits can be
represented with this simple model by the Norton's theorem. The
flux is coupled to the SQUID loop by a coil of inductance $L_{c}$.
We have introduced two partial inductances $L_{1}$ and $L_{2}$ \cite{Ruehli}
to model the linear inductance of the SQUID loop which are coupled
to the bias coil with mutual inductances $M_{1}$ and $M_{2}$. The
finely detailed circuit model in Fig.\,(\ref{fig:WTQ-IB}) that carefully
models the inductive network of the WTQ circuit and the flux bias
circuitry is crucial for avoiding ambiguities in the calculation of the
decoherence rates \cite{You-Koch,Riwar-DDV}. Our Born-Oppenheimer
treatment of the inductive network of the WTQ circuit, which reduces
the capacitance matrix to unity, finds the irrotational gauge discussed
in \cite{You-Koch} that removes the inconsistencies associated with
a gauge freedom.

\begin{figure}
	[tb]
	\begin{center}
		\includegraphics[
		width=\columnwidth 
		]%
		{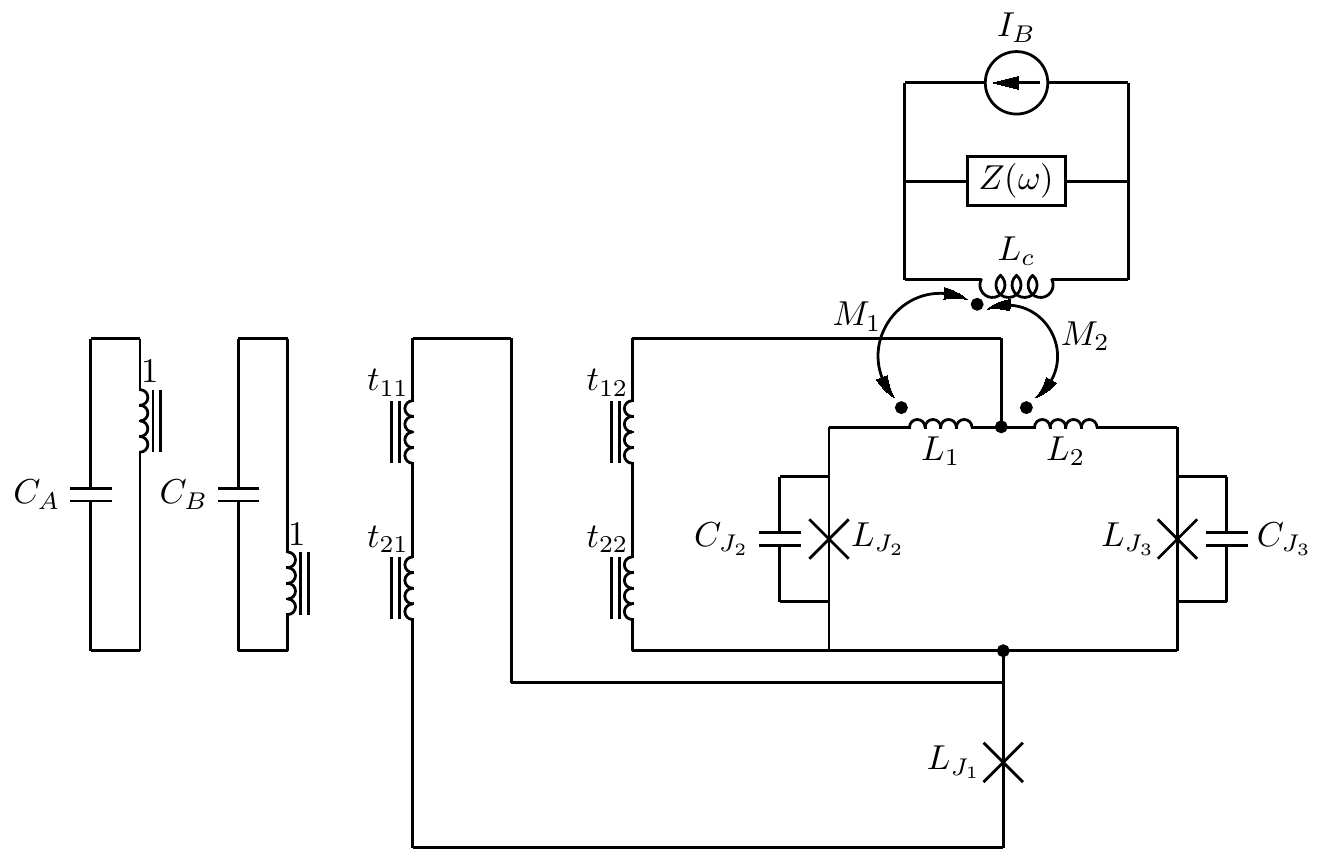}
		\caption{Multiport Belevitch transformer with a turns-ratio
			matrix $\mathbf{T}$ is introduced to reduce the number of capacitors
			by one. The capacitance values are $C_{A}=C_{1}+C_{J_{1}}+\frac{C_{c}C_{2}}{C_{2}+C_{c}}$
			and $C_{B}=C_{2}+C_{c}$. The turns-ratios are $t_{11}=1$, $t_{12}=\frac{C_{c}}{C_{2}+C_{c}}$, $t_{21}=0$ and $t_{22}=1$. Note that $C_{J_{1}}$ is added with $C_{1}$ and contributes to the value of the capacitance $C_{A}$.      
		}
		\label{fig:WTQ-T}
	\end{center}
\end{figure}

Before applying the formalism in \cite{BKD} we transform the capacitive
network in the original WTQ circuit in Fig.\,(\ref{fig:WTQ-IB}) with
the help of the multiport Belevitch transformer $\mathbf{T}$ as shown
in Fig.\,(\ref{fig:WTQ-T}). This transformation reduces the number
of capacitors by one and help us bring the circuit to a form treatable
in the formalism of \cite{BKD} by making the capacitance matrix of
the circuit diagonal:

\begin{equation}
\mathbf{C}_{0}=\left(\begin{array}{cccc}
C_{A} & 0 & 0 & 0\\
0 & C_{J_{2}} & 0 & 0\\
0 & 0 & C_{J_{3}} & 0\\
0 & 0 & 0 & C_{B}
\end{array}\right),
\end{equation}
where $C_{A}=C_{1}+C_{J_{1}}+\frac{C_{c}C_{2}}{C_{2}+C_{c}}$ and
$C_{B}=C_{2}+C_{c}$.

Also, since the number of degrees of freedom is given by the number
of capacitors in the minimum spanning tree of the circuit, we reduce
the number of degrees of freedom by one. The Belevitch transformer
turns-ratio matrix $\mathbf{T}$ is given by

\begin{align}
\mathbf{T} & =\left(\begin{array}{cc}
t_{11} & t_{12}\\
t_{21} & t_{22}
\end{array}\right)=\left(\begin{array}{cc}
1 & \frac{C_{c}}{C_{2}+C_{c}}\\
0 & 1
\end{array}\right),
\end{align}
which is obtained by a Cholesky Decomposition of the 2x2 impedance
matrix defined looking into the purely capacitive network consisting
of the capacitances $C_{1}$, $C_{2}$ and $C_{c}$.

The fundamental loop matrix defined in \cite{BKD} is given by:

\begin{equation}
\mathbf{F}_{CL}=\left(\begin{array}{cc}
-t_{12} & t_{12}\\
1 & 0\\
0 & -1\\
-t_{22} & t_{22}
\end{array}\right)=\left(\begin{array}{cc}
-\frac{C_{c}}{C_{2}+C_{c}} & \frac{C_{c}}{C_{2}+C_{c}}\\
1 & 0\\
0 & -1\\
-1 & 1
\end{array}\right).
\end{equation}

Note that the fundamental loop matrix of \cite{BKD} is generalized
to have real entries (turns ratios) other than $0$, $1$, and $-1$
\cite{MIQ}.

We use Eq.\,(62) of \cite{BKD} to calculate the inverse inductance
matrix $\mathbf{M}_{0}$ as

\begin{align}
\mathbf{M}_{0} & =\mathbf{F}_{CL}\tilde{\mathbf{L}}_{L}^{-1}\bar{\mathbf{L}}\mathbf{L}_{LL}^{-1}\mathbf{F}_{CL}^{T}.\label{eq:M0-definition}
\end{align}

To be able to calculate $\mathbf{M}_{0}$ we need to first introduce
some auxiliary matrices related to the inductive network in the circuit
that are defined in \cite{BKD}. The inductance matrix $\mathbf{L}_{t}$
is defined in Eq.\,(1) of \cite{BKD} as

\begin{align}
\mathbf{L}_{t} & =\left(\begin{array}{cc}
\mathbf{L} & \mathbf{L}_{LK}\\
\mathbf{L}_{LK}^{T} & \mathbf{L}_{K}
\end{array}\right)\\
& =\left(\begin{array}{ccc}
L_{1} & 0 & M_{1}\\
0 & L_{2} & M_{2}\\
M_{1} & M_{2} & L_{c}
\end{array}\right).
\end{align}

The inductance matrix $\mathbf{L}_{t}$ defined above is partitioned
according to the choice of the tree and chord inductors in the network
and the subscripts $L$ and $K$ denote chord and tree inductors,
respectively. Hence $L_{1}$ and $L_{2}$ are chord inductors, whereas
$L_{c}$ is a tree inductor. Eqs.\,(32) and (33) of \cite{BKD} define
the matrices $\mathbf{\bar{L}}$ and $\mathbf{\bar{L}}_{K}$ as

\begin{align}
\mathbf{\bar{L}} & =\mathbf{L}-\mathbf{L}_{LK}\mathbf{L}_{K}^{-1}\mathbf{L}_{LK}^{T}\\
& =\left(\begin{array}{cc}
L_{1}-\frac{M_{1}^{2}}{L_{c}} & -\frac{M_{1}M_{2}}{L_{c}}\\
-\frac{M_{1}M_{2}}{L_{c}} & L_{2}-\frac{M_{2}^{2}}{L_{c}}
\end{array}\right),
\end{align}

\begin{align}
\mathbf{\bar{L}}_{K} & =\mathbf{L}_{K}-\mathbf{L}_{LK}^{T}\mathbf{L}^{-1}\mathbf{L}_{LK}\\
& =L_{c}(1-k_{1}^{2}-k_{2}^{2}),
\end{align}

\noindent where we defined the inductive coupling coefficients $k_{1}$ and
$k_{2}$ as $k_{1}=\frac{M_{1}}{\sqrt{L_{c}L_{1}}}$ and $k_{2}=\frac{M_{2}}{\sqrt{L_{c}L_{2}}}$.
$\mathbf{\bar{L}}_{K}$ turns out to be a scalar since we have only
$L_{c}$ as a tree inductor in our circuit.

Next, we borrow some more definitions from \cite{BKD} that are used
in the definition of $\mathbf{M}_{0}$ in Eq.\,(\ref{eq:M0-definition})
above. Eq.\,(42) of \cite{BKD} reads

\begin{align}
\mathbf{\bar{F}}_{KL} & =\mathbf{F}_{KL}-\mathbf{L}_{K}^{-1}\mathbf{L}_{LK}^{T}\\
& =-\frac{1}{L_{c}}(\begin{array}{cc}
M_{1} & M_{2})\end{array},
\end{align}

\noindent where we used the fact that $\mathbf{F}_{KL}=(\begin{array}{cc}
0 & 0\end{array})$ since the only tree inductor $L_{c}$ does not belong to any of the
fundamental loops defined by the chord inductors $L_{1}$ and $L_{2}$.
Eq.\,(41) of \cite{BKD} defines another auxiliary inductance matrix
$\tilde{\mathbf{L}}_{K}$ as

\begin{align}
\tilde{\mathbf{L}}_{K} & =(\mathbf{1}_{K}-\mathbf{L}_{K}\bar{\mathbf{F}}_{KL}\mathbf{L}^{-1}\mathbf{L}_{LK}\bar{\mathbf{L}}_{K}^{-1})^{-1}\mathbf{L}_{K}\\
& =L_{c}(1-k_{1}^{2}-k_{2}^{2})\label{eq:Lk-tilde}
\end{align}

Again, similar to $\mathbf{\bar{L}}_{K}$ calculated above $\tilde{\mathbf{L}}_{K}$
turns out to be a scalar since we have only $L_{c}$ as a tree inductor
in the circuit. Next, we calculate two more matrices that appear in
the definition of $\mathbf{M}_{0}$ in Eq.\,(\ref{eq:M0-definition})
above. Eq.\,(51) of \cite{BKD} defines the matrix $\mathbf{L}_{LL}$
as 

\begin{align}
\mathbf{L}_{LL} & =\bar{\mathbf{L}}+\mathbf{F}_{KL}^{T}\tilde{\mathbf{L}}_{K}\mathbf{\bar{F}}_{KL}\\
& =\bar{\mathbf{L}},
\end{align}
since $\mathbf{F}_{KL}=(\begin{array}{cc}
0 & 0\end{array})$ as we noted above. Eq.\,(47) of \cite{BKD} defines the matrix $\tilde{\mathbf{L}}_{L}$
as

\begin{align}
\tilde{\mathbf{L}}_{L}^{-1} & =(\mathbf{1}_{L}+\mathbf{L}^{-1}\mathbf{L}_{LK}\mathbf{\bar{L}}_{K}^{-1}\tilde{\mathbf{L}}_{K}\bar{\mathbf{F}}_{KL})\bar{\mathbf{L}}^{-1}\\
& =\left(\begin{array}{cc}
1/L_{1} & 0\\
0 & 1/L_{2}
\end{array}\right).
\end{align}

Using the definitions above we can now evaluate the expression for
$\mathbf{M}_{0}$ given in Eq.\,(\ref{eq:M0-definition}) as

\begin{align}
\mathbf{M}_{0} & =\mathbf{F}_{CL}\begin{pmatrix}1/L_{1} & 0\\
0 & 1/L_{2}
\end{pmatrix}\mathbf{F}_{CL}^{T}\\
& =\begin{pmatrix}t_{12}^{2}\left(\frac{1}{L_{1}}+\frac{1}{L_{2}}\right) & -\frac{t_{12}}{L_{1}} & -\frac{t_{12}}{L_{2}} & t_{12}\left(\frac{1}{L_{1}}+\frac{1}{L_{2}}\right)\\
-\frac{t_{12}}{L_{1}} & \frac{1}{L_{1}} & 0 & -\frac{1}{L_{1}}\\
-\frac{t_{12}}{L_{2}} & 0 & \frac{1}{L_{2}} & -\frac{1}{L_{2}}\\
t_{12}\left(\frac{1}{L_{1}}+\frac{1}{L_{2}}\right) & -\frac{1}{L_{1}} & -\frac{1}{L_{2}} & \frac{1}{L_{1}}+\frac{1}{L_{2}}
\end{pmatrix}.
\end{align}

The coupling vector $\mathbf{S}_{0}$ to the current source $I_{B}$
is calculated using Eq.\,(66) of \cite{BKD} as 

\begin{align}
\mathbf{S}_{0} & =\mathbf{F}_{CB}-\mathbf{F}_{CL}(\mathbf{L}_{LL}^{-1})^{T}\bar{\mathbf{F}}_{KL}^{T}\tilde{\mathbf{L}}_{K}^{T}\mathbf{F}_{KB}\\
& =\begin{pmatrix}t_{12}\left((1-k_{2}^{2})\frac{M_{1}}{L_{1}}-(1-k_{1}^{2})\frac{M_{2}}{L_{2}}+k_{1}k_{2}\frac{(M_{2}-M_{1})}{\sqrt{L_{1}L_{2}}}\right)\\
-(1-k_{2}^{2})\frac{M_{1}}{L_{1}}-k_{1}k_{2}\frac{M_{2}}{\sqrt{L_{1}L_{2}}}\\
-(1-k_{1}^{2})\frac{M_{2}}{L_{2}}+k_{1}k_{2}\frac{M_{1}}{\sqrt{L_{1}L_{2}}}\\
(1-k_{2}^{2})\frac{M_{1}}{L_{1}}-(1-k_{1}^{2})\frac{M_{2}}{L_{2}}+k_{1}k_{2}\frac{(M_{2}-M_{1})}{\sqrt{L_{1}L_{2}}}
\end{pmatrix}\\
& \cong\begin{pmatrix}t_{12}\left(\frac{M_{1}}{L_{1}}-\frac{M_{2}}{L_{2}}+\frac{(M_{2}-M_{1})}{\sqrt{L_{1}L_{2}}}\right)\\
-\frac{M_{1}}{L_{1}}-\frac{M_{2}}{\sqrt{L_{1}L_{2}}}\\
-\frac{M_{2}}{L_{2}}+\frac{M_{1}}{\sqrt{L_{1}L_{2}}}\\
\frac{M_{1}}{L_{1}}-\frac{M_{2}}{L_{2}}+\frac{(M_{2}-M_{1})}{\sqrt{L_{1}L_{2}}}
\end{pmatrix},
\end{align}
using the auxiliary matrices introduced above and noting that $\mathbf{F}_{CB}=0$
since none of the capacitors in the WTQ circuit belongs to the fundamental
loop defined by the chord branch corresponding to the DC current source
$I_{B}$ and $\mathbf{F}_{KB}=-1$. In the last line above we used
the fact that the inductive coupling constants $k_{1}$, $k_{2}$
are small; i.e. $k_{1},k_{2}\ll1$ to simplify the expressions. We
note here that the coupling vector $\mathbf{\bar{m}}_{0}$ (defined
in Eq.\,(65) of \cite{BKD}) to the impedance $Z(\omega)$ is
given by

\begin{align}
\mathbf{\bar{m}}_{0} & =\mathbf{F}_{CZ}-\mathbf{F}_{CL}(\mathbf{L}_{LL}^{-1})^{T}\bar{\mathbf{F}}_{KL}^{T}\tilde{\mathbf{L}}_{K}^{T}\mathbf{F}_{KZ}\\
& =-\mathbf{S}_{0},
\end{align}
since $\mathbf{F}_{CZ}=0$ and $\mathbf{F}_{KZ}=1$. We will use $\mathbf{\bar{m}}_{0}$
in later sections to calculate the decoherence rates of the WTQ.

Hence, we can write the Hamiltonian of the WTQ circuit in the initial
frame as

\begin{equation}
\mathcal{H}=\frac{1}{2}\mathbf{Q}^{T}\mathbf{C}_{0}^{-1}\mathbf{Q}+\frac{1}{2}\mathbf{\Phi}^{T}\mathbf{M}_{0}\mathbf{\Phi}+\mathbf{\Phi}^{T}\mathbf{S}_{0}I_{B}-\sum\limits_{i=1}^{3}E_{J_{i}}\mathrm{cos}(\varphi_{J_{i}}),
\end{equation}
\noindent where $\mathbf{\Phi}=\frac{\Phi_{0}}{2\pi}\left(\varphi_{J_{1}},\varphi_{J_{2}},\varphi_{J_{3}},\varphi_{4}\right)$. Here, the first three coordinates are the phases across the junctions whereas the last coordinate $\varphi_{4}$ is the phase across the
capacitor $C_{B}$. $\mathbf{Q}$ is the vector of charge variables
canonically conjugate to the fluxes $\mathbf{\Phi}$.

To find the DC flux bias developed across each junction in the limit
of small loop inductances $L_{1}\rightarrow0$, $L_{2}\rightarrow0$
and to determine the coupling of the WTQ to the impedance $Z(\omega)$
we need to perform a few coordinate transformations:

\begin{align}
\mathbf{R}_{0} & =\begin{pmatrix}1 & 0 & 0 & 0\\
t_{12} & 1 & 0 & 1\\
t_{12} & 0 & 1 & 1\\
0 & 0 & 0 & 1
\end{pmatrix}\\
\mathbf{R}_{1} & =\begin{pmatrix}1 & 0 & 0 & 0\\
0 & 1 & 0 & 0\\
0 & 0 & 1 & 0\\
-t_{12}\frac{\left(C_{J_{2}}+C_{J_{3}}\right)}{C_{b}} & -\frac{C_{J_{2}}}{C_{b}} & -\frac{C_{J_{3}}}{C_{b}} & 1
\end{pmatrix},\\
\mathbf{R}_{2} & =\begin{pmatrix}1 & -\frac{C_{J_{2}}C_{c}}{C_{a}C_{b}} & -\frac{C_{J_{3}}C_{c}}{C_{a}C_{b}} & 0\\
0 & 1 & 0 & 0\\
0 & 0 & 1 & 0\\
0 & 0 & 0 & 1
\end{pmatrix},
\end{align}
\noindent where we defined

\begin{align}
C_{a} & =C_{1}'+\frac{C_{2}'C_{c}}{C_{2}'+C_{c}},\\
C_{b} & =C_{2}+C_{c}+C_{J_{2}}+C_{J_{3}},
\end{align}
with $C_{1}'=C_{1}+C_{J_{1}}$ and $C_{2}'=C_{2}+C_{J_{2}}+C_{J_{3}}$. Next we do a capacitance
re-scaling as required by the Born-Oppenheimer treatment of \cite{Brito}

\begin{equation}
\mathbf{A}=\begin{pmatrix}\sqrt{C_{a}} & 0 & 0 & 0\\
0 & \sqrt{C_{11}} & -\beta\sqrt{C_{22}} & 0\\
0 & 0 & \sqrt{C_{22}} & 0\\
0 & 0 & 0 & \sqrt{C_{b}}
\end{pmatrix}
\end{equation}
with 

\begin{align}
C_{11} & =C_{J_{2}}\frac{C_{\alpha,3}C_{\beta,3}}{C_{a}C_{b}},\\
C_{22} & =C_{J_{3}}\frac{C_{A}C_{B}}{C_{\alpha,3}C_{\beta,3}},\\
\beta & =\sqrt{\frac{C_{J_{2}}C_{J_{3}}}{C_{a}C_{b}}}\frac{\left(C_{1}'+C_{c}\right)}{\sqrt{C_{A}C_{B}}},
\end{align}
and

\begin{align}
C_{\alpha,3} & =C_{1}'+\frac{C_{c}\left(C_{2}+C_{J_{3}}\right)}{C_{2}+C_{c}+C_{J_{3}}},\\
C_{\beta,3} & =C_{2}+C_{c}+C_{J_{3}}.
\end{align}

All four coordinate transformations can be combined into one, i.e., $\mathbf{R}_{t}$, such that

\begin{equation}
\mathbf{R}_{t}=\mathbf{R}_{0}\mathbf{R}_{1}\mathbf{R}_{2}\mathbf{A}^{-1}.
\end{equation}
The initial capacitance matrix $\mathbf{C}_{0}$ is transformed into
the identity matrix by the total transformation $\mathbf{R}_{t}$
as

\begin{equation}
\mathbf{C}_{0}\rightarrow\mathbf{R}_{t}^{T}\mathbf{C}_{0}\mathbf{R}_{t}=\mathbf{1}.
\end{equation}

This ensures the choice of the irrotational gauge that removes the
inconsistencies in the decoherence rate calculations as discussed
in \cite{You-Koch}.

Initial flux coordinate vector $\mathbf{\Phi}$ is transformed into
the final capacitance re-scaled coordinate vector $\mathbf{f}$ by

\begin{equation}
\mathbf{\Phi}=\mathbf{R}_{t}\mathbf{f}.
\end{equation}

Phases across the junctions can be written in terms of the final coordinates
$\mathbf{f}=\left(f_{1,}f_{2},f_{3},f_{4}\right)^{T}$ as

\begin{align}
\varphi_{J_{1}} & =\alpha_{11}f_{1}+\alpha_{12}f_{2}+\alpha_{13}f_{3}+\alpha_{14}f_{4},\\
\varphi_{J_{2}} & =\alpha_{21}f_{1}+\alpha_{22}f_{2}+\alpha_{23}f_{3}+\alpha_{24}f_{4},\\
\varphi_{J_{3}} & =\alpha_{31}f_{1}+\alpha_{32}f_{2}+\alpha_{33}f_{3}+\alpha_{34}f_{4},
\end{align}
with

\begin{equation}
\alpha_{11}=\frac{1}{\sqrt{C_{a}}},\quad\alpha_{21}=\left(\frac{C_{c}}{C_{b}}\right)\frac{1}{\sqrt{C_{a}}},\quad\alpha_{31}=\left(\frac{C_{c}}{C_{b}}\right)\frac{1}{\sqrt{C_{a}}},
\end{equation}

\begin{equation}
\alpha_{14}=0,\quad\alpha_{24}=\frac{1}{\sqrt{C_{b}}},\quad\alpha_{34}=\frac{1}{\sqrt{C_{b}}},
\end{equation}

\begin{equation}
\alpha_{12}=-\frac{C_{c}\sqrt{C_{J_{2}}}}{\sqrt{C_{a}C_{b}C_{\alpha,3}C_{\beta,3}}},\quad\alpha_{13}=-\frac{C_{c}\sqrt{C_{J_{3}}}}{\sqrt{C_{A}C_{B}C_{\alpha,3}C_{\beta,3}}},
\end{equation}

\begin{equation}
\alpha_{22}=\frac{1}{\sqrt{C_{J_{2}}}}\sqrt{\frac{C_{\alpha,3}C_{\beta,3}}{C_{a}C_{b}}},\quad\alpha_{23}=0,
\end{equation}

\begin{equation}
\alpha_{32}=-\frac{\left(C_{1}'+C_{c}\right)\sqrt{C_{J_{2}}}}{\sqrt{C_{a}C_{b}C_{\alpha,3}C_{\beta,3}}},\quad\alpha_{33}=\frac{1}{\sqrt{C_{J_{3}}}}\sqrt{\frac{C_{A}C_{B}}{C_{\alpha,3}C_{\beta,3}}}.
\end{equation}

Hence, we can write the system Hamiltonian in the capacitance re-scaled
final coordinates $\mathbf{f}$ as

\begin{equation}
\mathcal{H}=\frac{1}{2}\mathbf{q}_{f}^{T}\mathbf{q}_{f}+U(\mathbf{f}),
\end{equation}

\noindent where $\mathbf{q}_{f}$ is the vector of momenta canonically conjugate
to the final coordinates $\mathbf{f}$ and the potential $U(\mathbf{f})$
is given by

\begin{equation}
U(\mathbf{f})=\frac{1}{2}\mathbf{f}^{T}\mathbf{R}_{t}^{T}\mathbf{M}_{0}\mathbf{R}_{t}\mathbf{f}+\mathbf{f}^{T}\mathbf{R}_{t}^{T}\mathbf{S}_{0}I_{B}-\sum\limits_{i=1}^{3}E_{J_{i}}\mathrm{cos}(\varphi_{J_{i}}).\label{eq:U-prime}
\end{equation}

\subsection*{Born-Oppenheimer Approximation}

We perform the Born-Oppenheimer approximation following \cite{Brito}.
The coordinates $f_{2}$ and $f_{3}$ are fast coordinates denoted
by the vector $\mathrm{\mathbf{f}_{\bot}}=(f_{2},\,f_{3})$ and will
be eliminated. Since the potential seen is very steep along the direction
of the fast coordinates it can be approximated with a harmonic potential
as given by the Eq.\,(14) in \cite{Brito}

\begin{equation}
U(\mathbf{f})\approx V(\mathbf{f}_{\Vert})+\underset{i}{\sum}a_{i}(\mathbf{f}_{\bot})_{i}+\underset{i,j}{\sum}b_{ij}(\mathbf{f}_{\bot})_{i}(\mathbf{f}_{\bot})_{j},\label{eq:BO-fast-potential}
\end{equation}
\noindent where $\mathrm{\mathbf{f}_{\Vert}}=(f_{1},\,f_{4})$ is the vector
holding the slow coordinates and the matrix $\mathbf{b}=[b_{ij}]$
is the 2x2 sector (corresponding to the coordinates $f_{2}$ and $f_{3}$)
of the transformed $\mathbf{M}_{0}$ matrix in the final frame; i.e.
of the matrix $\mathbf{R}_{t}^{T}\mathbf{M}_{0}\mathbf{R}_{t}$:

\begin{equation}
\mathbf{b}=\left(\begin{array}{cc}
\frac{1}{\alpha_{22}^{2}C_{J_{2}}^{2}L_{1}} & \frac{1}{C_{23}L_{1}}\\
\frac{1}{C_{23}L_{1}} & \frac{1}{\alpha_{33}^{2}C_{J_{3}}^{2}L_{2}}+\frac{(C_{1}'+C_{c})^{2}C_{J_{3}}}{C_{A}C_{B}C_{\alpha,3}C_{\beta,3}L_{1}}
\end{array}\right),
\end{equation}
\noindent where we defined $C_{23}^{-1}=\sqrt{\frac{C_{J_{3}}C_{a}C_{b}}{C_{J_{2}}C_{A}C_{B}}}\frac{(C_{1}'+C_{c})}{C_{\alpha,3}C_{\beta,3}L_{1}}$.
The vector $\mathbf{a}=(a_{1},\,a_{2})$ in Eq.\,(\ref{eq:BO-fast-potential})
holds the entries (corresponding to the fast coordinates $f_{2}$
and $f_{3})$ of the transformed $\mathbf{S}_{0}$ vector in the final
frame; i.e. of the vector $\mathbf{R}_{t}^{T}\mathbf{S}_{0}$ (with
the scale factor $I_{B}/\varphi_0$):

\begin{equation}
\mathbf{a}=\left(\begin{array}{c}
-\frac{M_{1}}{\alpha_{22}C_{J_{2}}L_{1}}\\
-\frac{(C_{1}'+C_{c})\sqrt{C_{J_{3}}}M_{1}}{\sqrt{C_{A}C_{B}C_{\alpha,3}C_{\beta,3}}L_{1}}+\frac{M_{2}}{\alpha_{33}C_{J_{3}}L_{2}}
\end{array}\right)\left(\frac{I_{B}}{\varphi_{0}}\right),
\end{equation}

\noindent where $\varphi_0\equiv\Phi_{0}/2\pi$ is the reduced flux quantum.
 
The potential is centered around

\begin{align}
\left(\mathbf{f}_{\perp}^{\rm{min}}\right)_{0} & =\nonumber \\
&\begin{pmatrix}\alpha_{22}C_{J_{2}}\left(M_{1}+\frac{\left(C_{1}'+C_{c}\right)C_{J_{3}}}{C_{\alpha,3}C_{\beta,3}}M_{2}\right)\\
-\alpha_{33}C_{J_{3}}M_{2}
\end{pmatrix}\left(\frac{I_{B}}{\varphi_0}\right)\label{eq:fmin-0}
\end{align}

\noindent for the fast coordinates (here, we dropped terms quadratic in the small
inductive coupling coefficients $k_{1}$, $k_{2}$). This is the DC
component of the fast coordinates that does not depend on the slow
coordinates. We used the Eq.\,(16) of \cite{Brito} to calculate $\left(\mathbf{f}_{\perp}^{min}\right)_{0}$:

\begin{equation}
\left(\mathbf{f}_{\perp}^{\rm{min}}\right)_{0}=-\frac{1}{2}\mathbf{b}^{-1}\mathbf{a}.
\end{equation}

Hence, the reduced DC flux bias in the phases of the junctions is

\begin{align}
\mathbf{\varphi}_{x}^{(0)} & =\left(\begin{array}{c}
\varphi_{x,1}^{(0)}\\
\varphi_{x,2}^{(0)}\\
\varphi_{x,3}^{(0)}
\end{array}\right)=\left(\begin{array}{cc}
\alpha_{12} & \alpha_{13}\\
\alpha_{22} & \alpha_{23}\\
\alpha_{32} & \alpha_{33}
\end{array}\right)\left(\mathbf{f}_{\perp}^{min}\right)_{0}\\
& =\begin{pmatrix}\frac{C_{c}\left(C_{J_{3}}M_{2}-C_{J_{2}}M_{1}\right)}{C_{a}C_{b}}\\
\frac{C_{\alpha,3}C_{\beta,3}}{C_{a}C_{b}}M_{1}+\frac{\left(C_{1}'+C_{c}\right)C_{J_{3}}}{C_{a}C_{b}}M_{2}\\
-\frac{\left(C_{1}'+C_{c}\right)C_{J_{2}}}{C_{a}C_{b}}M_{1}-\frac{C_{\alpha,2}C_{\beta,2}}{C_{a}C_{b}}M_{2}
\end{pmatrix}\left(\frac{I_{B}}{\varphi_{0}}\right),
\end{align}
\noindent where we defined

\begin{align}
C_{\alpha,2} & =C_{1}'+\frac{C_{c}(C_{2}+C_{J_{2}})}{C_{2}+C_{c}+C_{J_{2}}},\\
C_{\beta,2} & =C_{2}+C_{c}+C_{J_{2}}.
\end{align}

We note here that $\varphi_{x}=\varphi_{x,2}^{(0)}-\varphi_{x,3}^{(0)}=\left(M_{1}+M_{2}\right)\left(\frac{I_{B}}{\varphi_{0}}\right)$, as expected.

We perform the Born-Oppenheimer approximation by expanding the cosine
potentials around the DC flux biases in Eq.\,(\ref{eq:fmin-0}) for
the fast coordinates to obtain the following effective Hamiltonian:

\begin{equation}
\mathcal{H}=\frac{1}{2}\mathbf{q}^{T}\mathcal{C}^{-1}\mathbf{q}-E_{J_{1}}\mathrm{cos}\left(\varphi_{1}+\varphi_{x,1}^{(0)}\right)-E_{2}\mathrm{cos}\left(\varphi_{2}+\varphi_{x}^{(2)}\right),\label{eq:2-deg-free-H}
\end{equation}
\noindent where

\begin{equation}
\mathcal{C}=\left(\begin{array}{cc}
C_{1}'+C_{c} & -C_{c}\\
-C_{c} & C_{2}'+C_{c}
\end{array}\right),
\end{equation}

\begin{align}
E_{2} & =\left(E_{J_{2}}+E_{J_{3}}\right)\mathrm{cos}\left( \dfrac{\varphi_{x}}{2}\right) \sqrt{1+d^{2}\mathrm{tan}^{2}\left( \dfrac{\varphi_{x}}{2}\right) },\\
\varphi_{x}^{(2)} & =\frac{\left(\varphi_{x,2}^{(0)}+\varphi_{x,3}^{(0)}\right)}{2}-\mathrm{tan}^{-1}\left(d\mathrm{tan}\left( \dfrac{\varphi_{x}}{2}\right) \right),
\end{align}
with $d=\left(E_{J_{3}}-E_{J_{2}}\right)/\left(E_{J_{2}}+E_{J_{3}}\right)$,
$\varphi_{1}=f_{1}/\sqrt{C_{a}}$, $\varphi_{2}=f_{2}/\sqrt{C_{b}}$, $E_{J_{1}}=\varphi_0I_{J1}$, $E_{J_{2}}=\varphi_0I_{J2}$, and $E_{J_{3}}=\varphi_0I_{J3}$.

Eq.\,(\ref{eq:2-deg-free-H}) is the Hamiltonian corresponding to two
transmon qubits with Josephson energies $E_{J_{1}}$ and $E_{2}$
and shunting capacitances $C_{1}'$ and $C_{2}'$
coupled electrostatically with capacitance $C_{c}$.

\section{WTQ frequency and anharmonicity}
At this point we can calculate the qubit frequency $\omega_{q}$ using
formulas derived in \cite{Z-paper}

\begin{equation}
\omega_{q}=\omega_{1}\sqrt{1-\frac{r^{2}\omega_{1}^{2}}{\omega_{2}^{2}-(1-r^{2})\omega_{1}^{2}}},\label{eq:wq}
\end{equation}
\noindent where $\omega_{1}$ is the bare frequency of the qubit mode given
by \cite{Z-paper}

\begin{equation}
\omega_{1}=\omega_{J_{1}}-\frac{E_{C_{1}}/\hbar}{1-E_{C_{1}}/(\hbar\omega_{J_{1}})},
\end{equation}
with 

\begin{equation}
E_{C_{1}}=\frac{e^{2}}{2\left(C_{1}'+C_{c}\right)},
\end{equation}
and

\begin{equation}
\omega_{J_{1}}=\frac{1}{\sqrt{L_{J_{1}}\left(C_{1}'+C_{c}\right)}}.
\end{equation}

The coupling coefficient $r$ in Eq.\,(\ref{eq:wq}) between the modes
of the WTQ is defined by

\begin{equation}
r=\frac{C_{c}}{\sqrt{C_{1}'C_{2}'+C_{c}\left(C_{1}'+C_{2}'\right)}}.
\end{equation}

The charging energy $E_{C_{2}}$ for the high-frequency SQUID mode is

\begin{equation}
E_{C_{2}}=\frac{e^{2}\left(1+r^{2}\right)}{2\left(C_{2}'+C_{c}\right)}.
\end{equation}
Bare frequency $\omega_{2}$ of the SQUID mode is given by

\begin{equation}
\omega_{2}=\omega_{J_{2}}-\frac{E_{C_{2}}/\hbar}{1-E_{C_{2}}/\left(\hbar\omega_{J_{2}}\right)},
\end{equation}
with 

\begin{equation}
\omega_{J_{2}}=\frac{1}{\sqrt{L_{J_{S}}\frac{\left(C_{2}'+C_{c}\right)}{\left(1+r^{2}\right)}}},
\end{equation}
\noindent where we defined the effective SQUID inductance $L_{J_{S}}$

\begin{equation}
L_{J_{S}}=\left(\left(\dfrac{1}{L_{J_{2}}}+\dfrac{1}{L_{J_{3}}}\right)\left| \mathrm{cos}\left( \dfrac{\varphi_{x}}{2}\right)\right|  \sqrt{1+d^{2}\mathrm{tan^{2}}\left(\dfrac{\varphi_{x}}{2}\right) }\right)^{-1}.
\end{equation}

Anharmonicity $\alpha$ of the qubit mode can be estimated by
\cite{Z-paper}

\begin{equation}
\alpha=-E_{C_{1}}\left(\frac{\omega_{J_{1}}}{\omega_{1}}\right)^{2}\left(1-\frac{r^{2}\omega_{1}^{2}}{\omega_{2}^{2}-(1-r^{2})\omega_{1}^{2}}\right)^{3}.\label{eq:Anharmonicity-formula}
\end{equation}

In Fig.\,\ref{GeneralWTQparams}(a),(b) we plot the
transition frequency $f_{01}=f_q=\omega_q/2\pi$ and the anharmonicity $\alpha$ of a WTQ as a function of the normalized applied flux. The device parameters in this example are chosen to yield a WTQ with $f^{\rm{max}}_{01}$ and $\alpha$ of about $5$ GHz and $300$ MHz, respectively, and frequency tunability $\delta=50$ MHz. In the calculation, we use the analytical
formulas of Eqs.\,(\ref{eq:wq}) and (\ref{eq:Anharmonicity-formula}), which we plot as red dashed curves and compare them to the results obtained using the exact diagonalization
of the qubit Hamiltonian specified in Eq.\,(\ref{eq:2-deg-free-H}) in the charge basis
(Bloch-wave basis), which we plot as blue solid curves. As seen in Fig.\,\ref{GeneralWTQparams}(b), the WTQ anharmonicity varies with the applied flux. But the variation is relatively small of about $17$ MHz in this example, between the minimum and maximum sweet spots (based on the exact calculation). Similarly, we plot in Fig.\,\ref{GeneralWTQparams}(c), using blue solid and red dashed curves, the exact and analytical solutions for the high-frequency mode of the qubit $f_{10}$ versus the normalized applied flux. The slight bending in the exact diagonalization curve (the solid blue), seen around $14$ GHz, is due to the crossing of $f_{10}$, the first excited level of the SQUID oscillator, and the third level of the qubit.

\section{Calculation of the Decoherence Rates}

We will now employ Fermi Golden-Rule type formulas in Eqs.\,(10-11)
of \cite{Brito} to calculate the relaxation and dephasing rates of
the WTQ:

\begin{figure*}
	[tb]
	\begin{center}
		\includegraphics[
		width=1.8\columnwidth 
		]%
		{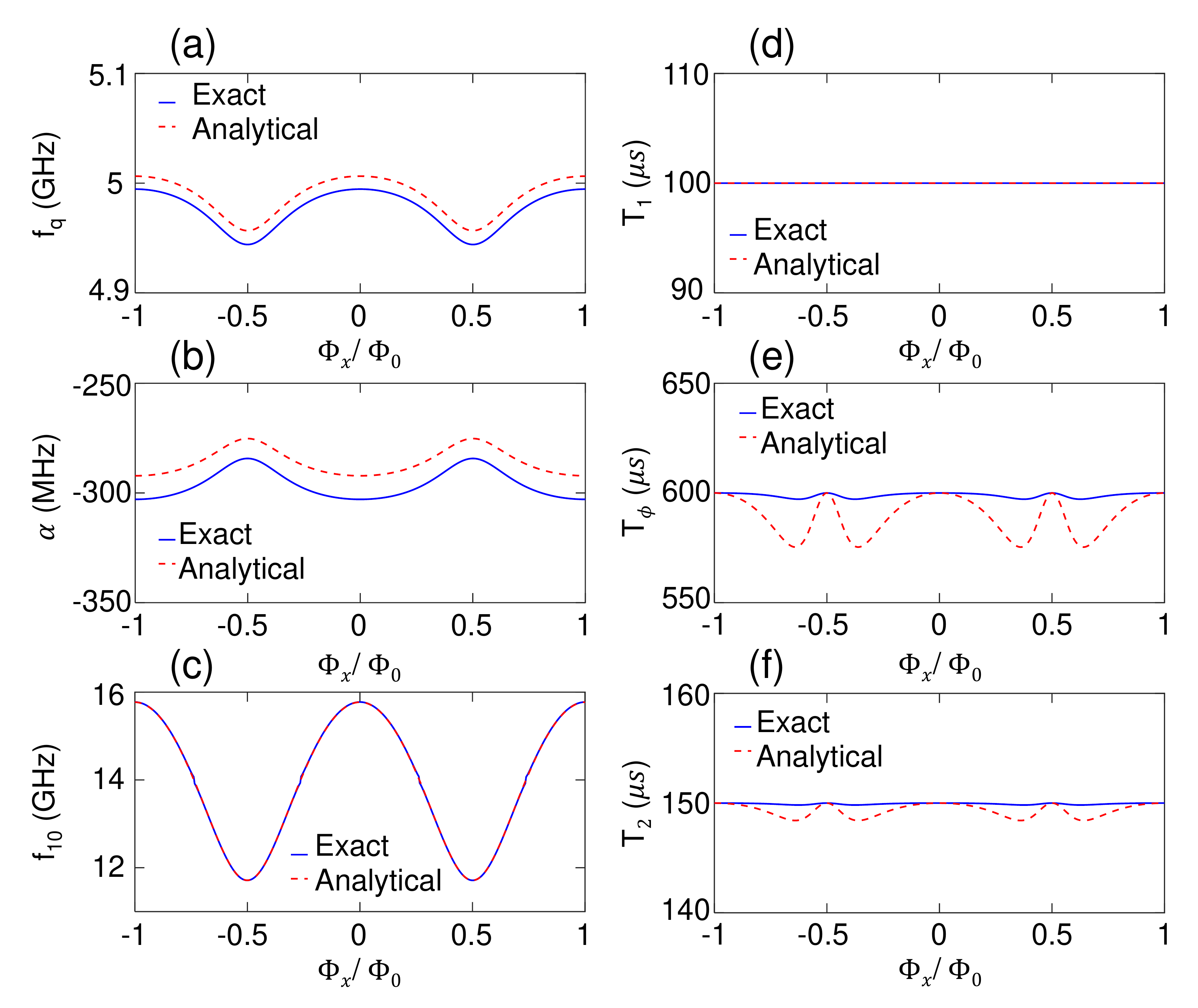}
		\caption{A WTQ example. (a) Qubit  frequency and (b) anharmonicity of the WTQ as a function of the normalized flux threading the SQUID loop. In this example, the WTQ has a maximum frequency and anharmonicity of about $5$ GHz and $300$ MHz, respectively and exhibit a frequency tunability $\delta=50$ MHz. Blue curves are obtained by the diagonalization of the Hamiltonian
			in Eq.\,(\ref{eq:2-deg-free-H}) in the charge basis. Red dashed curves are
			calculated using the analytical formulas in Eqs.\,(\ref{eq:wq}) and
			(\ref{eq:Anharmonicity-formula}). (c) The high-frequency mode of the WTQ $f_{10}$ versus normalized flux. The WTQ circuit parameters employed in this example are $I_{c1}=I_{c2}=26$ nA, $I_{c3}=\alpha_JI_{c2}$, $\alpha_J=3.5$, $C_{1}'=50$ fF, $C_{2}'=20$ fF, $C_{c}=20$ fF, where $C_{J_{1}}=C_{J_{2}}=1$ fF and $C_{J_{3}}=\alpha_JC_{J_{2}}$. In plots (d), (e), (f), we evaluate the dependence of the WTQ coherence times, i.e., $T_1$, $T_{\phi}$, and $T_2$ on the applied flux. In this calculation, we assume $T=0.02$ K and $Z(\omega)=R=0.1$ Ohm and that the maximum values for $T_1$ and $T_2=1.5T_1$, i.e., $100$ $\mu$s and $150$ $\mu$s respectively, are set by loss or noise sources that are independent of flux or the flux biasing circuit. The $T_2$ curves in (f) are calculated using the relation $1/T_2=1/2T_1+1/T_{\phi}$, which also sets the maximum for $T_{\phi}$ in (e). The analytical calculation of $T_1$ and $T_{\phi}$ uses the matrix element results of Eqs.\,(\ref{TransitionEqFirst}-\ref{TransitionEqLast}). The exact calculation employs Eqs.\,(\ref{eq:T1}), (\ref{eq:Tphi}) with a diagonalization of the Hamiltonian in Eq.\,(\ref{eq:2-deg-free-H}).
		}
		\label{GeneralWTQparams}
	\end{center}
\end{figure*}

\begin{figure*}
	[tb]
	\begin{center}
		\includegraphics[
		width=1.8\columnwidth 
		]%
		{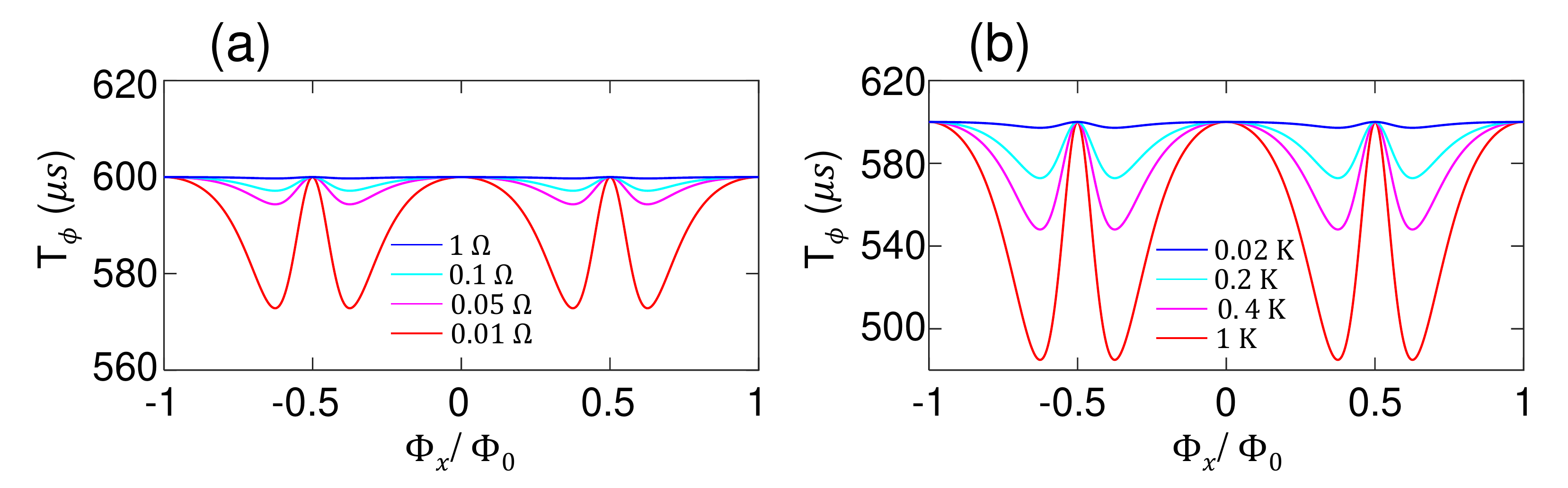}
		\caption{Dependence of WTQ dephasing time $T_{\phi}$ on the normalized flux threading the SQUID loop for varying parameter $R$ (a) and $T$ (b) of the flux biasing circuit, where $Z(\omega)=R$ is assumed. In (a) and (b) the values of $T$ and $R$ are set to $0.02$ K and $0.1$ Ohm, respectively. The WTQ circuit parameters employed in this calculation are the same as Fig.\,\ref{GeneralWTQparams}. The curves in both plots are obtained by the diagonalization of the Hamiltonian in Eq.\,(\ref{eq:2-deg-free-H}) in the charge basis along with Eqs.\,(\ref{eq:T1}), (\ref{eq:Tphi}).    
		}
		\label{GeneralWTQdephasing}
	\end{center}
\end{figure*}

\begin{align}
\frac{1}{T_{1}} & =\frac{4}{\hbar}\left|\left\langle 0\right|\mathbf{\bar{m}}^{T}\mathbf{f}\left|1\right\rangle \right|^{2}J\left(\omega_{01}\right)\mathrm{coth}\left(\frac{\hbar\omega_{01}}{2k_{B}T}\right),\label{eq:T1}\\
\frac{1}{T_{\phi}} & =\frac{1}{\hbar}\left|\left\langle 0\right|\mathbf{\bar{m}}^{T}\mathbf{f}\left|0\right\rangle -\left\langle 1\right|\mathbf{\bar{m}}^{T}\mathbf{f}\left|1\right\rangle \right|^{2}\left.\frac{J(\omega)}{\hbar\omega}\right|_{\omega\rightarrow0}2k_{B}T,\label{eq:Tphi}
\end{align}
\noindent where $\mathbf{\bar{m}}$ is the vector giving the coupling to the
impedance in the final frame, that is

\begin{align}
\mathbf{\bar{m}} & =\mathbf{R}_{t}^{T}\mathbf{\bar{m}_{0}}\nonumber\\
& =\left(\begin{array}{c}
0\\
\frac{C_{a}C_{b}M_{1}}{\sqrt{C_{J_{2}}}C_{\alpha,3}C_{\beta,3}L_{1}}\\
\frac{(C_{1}'+C_{c})\sqrt{C_{J_{3}}}M_{1}}{\sqrt{C_{A}C_{B}C_{\alpha,3}C_{\beta,3}}L_{1}}-\sqrt{\frac{C_{\alpha,3}C_{\beta,3}}{C_{A}C_{B}C_{J_{3}}}}\frac{M_{2}}{L_{2}}\\
0
\end{array}\right).\label{eq:mbar}
\end{align}

Hence, we see that only the fast coordinates ($f_{2}$ and $f_{3}$)
are coupled to the decoherence source which is the impedance $Z(\omega)$.

The spectral density $J(\omega)$ of the bath due to the impedance
$Z(\omega)$ is calculated using Eq.\,(93) of \cite{BKD} (up to the
scale factors)

\begin{equation}
J(\omega)=\mathrm{Im}\left[K(\omega)\right]=\frac{\omega\mathrm{Re}\left[Z(\omega)\right]}{\omega^{2}L_{c}^{2}+\left|Z\right|^{2}+2\omega L_{c}\mathrm{Im}\left[Z(\omega)\right]},\label{eq:Jw}
\end{equation}
\noindent where $K(\omega)$ is the kernel function of the bath given in Eq.\,(73)
of \cite{BKD} as

\begin{equation}
K(\omega)=\bar{\mathbf{L}}_{Z}^{-1}(\omega).\label{eq:Kw}
\end{equation}

$\bar{\mathbf{L}}_{Z}$ is defined in Eq.\,(58) of \cite{BKD} as

\begin{equation}
\bar{\mathbf{L}}_{Z}=\mathbf{L}_{ZZ}-\mathbf{L}_{ZL}\mathbf{L}_{LL}^{-1}\mathbf{L}_{LZ},\label{eq:LZ-bar}
\end{equation}
\noindent where $\mathbf{L}_{ZZ}$ is given in Eq.\,(52) of \cite{BKD} as

\begin{align}
\mathbf{L}_{ZZ} & =\mathbf{L}_{Z}+\mathbf{F}_{KZ}^{T}\tilde{\mathbf{L}}_{K}\mathbf{F}_{KZ}\\
& =\mathbf{L}_{Z}+\tilde{\mathbf{L}}_{K}\\
& =\frac{Z(\omega)}{i\omega}+L_{c}(1-k_{1}^{2}-k_{2}^{2}),
\end{align}
\noindent where in the second line above we used the fact that $\mathbf{F}_{KZ}=1$
and in the third line we used Eq.\,(\ref{eq:Lk-tilde}). $\mathbf{L}_{ZL}$
is given by Eq.\,(53) of \cite{BKD} as

\begin{equation}
\mathbf{L}_{ZL}=\mathbf{F}_{KL}^{T}\tilde{\mathbf{L}}_{K}\mathbf{F}_{KZ}=0,
\end{equation}
since $\mathbf{F}_{KL}=0$. Hence, $\bar{\mathbf{L}}_{Z}$ in Eq.\,(\ref{eq:LZ-bar})
above is given by

\begin{align}
\bar{\mathbf{L}}_{Z} & =\mathbf{L}_{ZZ}\\
& =\frac{Z(\omega)}{i\omega}+L_{c}(1-k_{1}^{2}-k_{2}^{2}).
\end{align}
Now Eq.\,(\ref{eq:Kw}) for $K(\omega)$ reads

\begin{align}
K(\omega) & =\bar{\mathbf{L}}_{Z}^{-1}(\omega)\\
& =\frac{i\omega}{Z(\omega)+i\omega L_{c}(1-k_{1}^{2}-k_{2}^{2})}.\label{K}
\end{align}

Furthermore, calculating $\mathrm{Im}\left[K(\omega)\right]$ using Eq.\,(\ref{K}) yields the result of Eq.\,(\ref{eq:Jw}) since $k_{1},k_{2}\ll1$.

The dependence of the argument $\mathbf{\bar{m}}^{T}\mathbf{f}$ of
the matrix elements on the slow coordinates $\varphi_{1}$ and $\varphi_{2}$
can be calculated using Eqs.\,(16-20) of \cite{Brito}. Eq.\,(16) of
\cite{Brito} reads

\begin{align}
\left(\mathbf{f}_{\perp}^{min}\right)(\mathrm{\mathbf{f}_{\parallel})} & =-\frac{1}{2}\mathbf{b}^{-1}\mathbf{a}(\mathrm{\mathbf{f}_{\parallel})},
\end{align}
\noindent where $\left(\mathbf{f}_{\perp}^{min}\right)(\mathrm{\mathbf{f}_{\parallel})}$
gives the dependence of the fast coordinates on the slow coordinates
with 

\begin{equation}
\mathbf{a}(\mathrm{\mathbf{f}_{\parallel})}=\left(\begin{array}{ccc}
\alpha_{12} & \alpha_{22} & \alpha_{32}\\
\alpha_{13} & \alpha_{23} & \alpha_{33}
\end{array}\right)\left(\begin{array}{c}
\mathrm{sin}\left(\varphi_{J_{1}}\right)/L_{J_{1}}\\
\mathrm{sin}\left(\varphi_{J_{2}}\right)/L_{J_{2}}\\
\mathrm{sin}\left(\varphi_{J_{3}}\right)/L_{J_{3}}
\end{array}\right).
\end{equation}

Hence,

\begin{align}
\mathbf{\bar{m}}^{T}\mathbf{f} & =\mathbf{\bar{m}_{\perp}}^{T}\left(\mathbf{f}_{\perp}^{min}\right)(\mathrm{\mathbf{f}_{\parallel})}\label{eq:mfT}\\
& =A_{s}^{(1)}\mathrm{sin}\left(\varphi_{J_{1}}\right)+A_{s}^{(2)}\mathrm{sin}\left(\varphi_{J_{2}}\right)+A_{s}^{(3)}\mathrm{sin}\left(\varphi_{J_{3}}\right),
\end{align}

\noindent where $\mathbf{\bar{m}_{\perp}}$ is the sub-vector of $\mathbf{\bar{m}}$
corresponding to the fast coordinates and

\begin{equation}
\mathbf{A}_{s}=\left(\begin{array}{c}
A_{s}^{(1)}\\
A_{s}^{(2)}\\
A_{s}^{(3)}
\end{array}\right)=\begin{pmatrix}\frac{C_{c}\left(C_{J_{2}}M_{1}-C_{J_{3}}M_{2}\right)}{C_{a}C_{b}L_{J_{1}}}\\
-\frac{C_{\alpha,3}C_{\beta,3}M_{1}+(C_{1}'+C_{c})C_{J_{3}}M_{2}}{C_{a}C_{b}L_{J_{2}}}\\
\frac{\left(C_{1}'+C_{c}\right)C_{J_{2}}M_{1}+C_{\alpha,2}C_{\beta,2}M_{2}}{C_{a}C_{b}L_{J_{3}}}
\end{pmatrix}.
\end{equation}

Note that we dropped the DC flux bias terms in the expression in Eq.\,(\ref{eq:mfT}) above. The last two terms in the above expression
can again be combined in a single sine function as:

\begin{align}
\mathbf{\bar{m}}^{T}\mathbf{f} &
=A_{s}^{(1)}\mathrm{sin}\left(\varphi_{1}+\varphi_{x,1}^{(0)}\right) \nonumber\\ 
&+A\mathrm{cos}\left( \dfrac{\varphi_{x}}{2}\right) \sqrt{1+d_{s}^{2}\mathrm{tan}^{2}\left( \dfrac{\varphi_{x}}{2}\right)} \mathrm{sin}\left(\varphi_{2}+\varphi_{x}^{(s)}\right),
\end{align}

\noindent where $A=A_{s}^{(2)}+A_{s}^{(3)}$, $d_{s}=\left( A_{s}^{(2)}-A_{s}^{(3)}\right) /\left( A_{s}^{(2)}+A_{s}^{(3)}\right) $
and $\varphi_{x}^{(s)}=\left(\varphi_{x,2}^{(0)}+\varphi_{x,3}^{(0)}\right)/2+\mathrm{tan}^{-1}\left(d_{s}\mathrm{tan}\left(\varphi_{x}/2\right) \right)$.

Matrix elements can be evaluated by assuming harmonic wavefunctions
for the main qubit mode and the SQUID mode. Since the coupling between
the two modes is dispersive, we can write the excited level $\widetilde{\left|1\right\rangle }$
of the qubit as:

\begin{equation}
\widetilde{\left|1\right\rangle }\approx\left(1-\frac{\epsilon^{2}}{2}\right)\left|10\right\rangle -\epsilon\left|01\right\rangle, \label{eq:Excited-state}
\end{equation}
\noindent where $\epsilon=\frac{J_{12}}{\Delta}$. Here the excitation on the left
under the ket corresponds to the qubit mode, whereas the right excitation
corresponds to the SQUID mode and $\Delta=\omega_{2}-\omega_{1}$
is the detuning between the qubit and SQUID modes. The coupling rate
$J_{12}$ between the qubit and SQUID mode is given by

\begin{equation}
J_{12}=\frac{1}{2\sqrt{Z_{1}Z_{2}}}\frac{C_{c}}{C_{1}'C_{2}'+C_{c}(C_{1}'+C_{2}')},
\end{equation}
\noindent where $Z_{1}$ is the characteristic impedance of the qubit mode and
$Z_{2}$ is the characteristic impedance of the SQUID mode, which are
given by

\begin{align*}
Z_{1} & =\frac{1}{\omega_{1}(C_{1}'+C_{c})},\\
Z_{2} & =\sqrt{\frac{L_{J_{S}}}{\left(C_{2}'+C_{c}\right)\left(1-\frac{2E_{C_{2}}}{\hbar\omega_{2}}\right)}}.
\end{align*}

Using symmetry arguments one can show that

\begin{align}
\widetilde{\left\langle 0\right|}\mathrm{sin}\left(\varphi_{1}+\varphi_{x}^{(1)}\right)\widetilde{\left|0\right\rangle } & =0, \label{TransitionEqFirst} \\ 
\widetilde{\left\langle 1\right|}\mathrm{sin}\left(\varphi_{1}+\varphi_{x}^{(1)}\right)\widetilde{\left|1\right\rangle } & =0.  
\end{align}

Using Eq.\,(\ref{eq:Excited-state}), we get

\begin{align}
\widetilde{\left\langle 1\right|}\mathrm{sin}\left(\varphi_{2}+\varphi_{x}^{(s)}\right)\widetilde{\left|1\right\rangle }&
=\left( 1-\frac{\epsilon^{2}}{2}\right) ^{2}\left\langle 10\right|\mathrm{sin}\left(\varphi_{2}+\varphi_{x}^{(s)}\right)\left|10\right\rangle \nonumber\\
&+\epsilon^{2}\left\langle 01\right|\mathrm{sin}\left(\varphi_{2}+\varphi_{x}^{(s)}\right)\left|01\right\rangle. \nonumber\\
\end{align}

Hence,

\begin{align}
\left\langle 0\right|\mathbf{\bar{m}}^{T}\mathbf{f}\left|0\right\rangle -\left\langle 1\right|\mathbf{\bar{m}}^{T}\mathbf{f}\left|1\right\rangle  & \approx\epsilon^{2}\left\langle 10\right|\mathrm{sin}\left(\varphi_{2}+\varphi_{x}^{(s)}\right)\left|10\right\rangle \nonumber\\
&-\epsilon^{2}\left\langle 01\right|\mathrm{sin}\left(\varphi_{2}+\varphi_{x}^{(s)}\right)\left|01\right\rangle \nonumber\\
& =\epsilon^{2}\eta_{2}^{2}\mathrm{exp}\left(-\dfrac{\eta_{2}^{2}}{2}\right)\mathrm{sin}\left(\varphi_{\Delta}\right),
\end{align}
\noindent where 

\begin{equation}
\varphi_{\Delta}=\mathrm{tan}^{-1}\left(d\mathrm{tan}\left( \dfrac{\varphi_{x}}{2}\right) \right)+\mathrm{tan}^{-1}\left(d_{s}\mathrm{tan}\left( \dfrac{\varphi_{x}}{2}\right)\right),
\end{equation}

\begin{equation}
\eta_{2}^{2}=4\pi\left(\frac{Z_{2}}{R_{Q}}\right),
\end{equation}

\noindent and $R_{Q}=\frac{h}{e^{2}}\approx25813\Omega$ is the resistance quantum. 

The matrix elements in the $T_{1}$ expression in Eq.\,(\ref{eq:T1})
can be calculated as:

\begin{align}
\widetilde{\left\langle 0\right|}\mathrm{sin}\left(\varphi_{1}+\varphi_{x}^{(1)}\right)\widetilde{\left|1\right\rangle } & =\left\langle 00\right|\mathrm{sin}\left(\varphi_{1}+\varphi_{x}^{(1)}\right)\left|10\right\rangle \nonumber\\
& =\eta_{1}\mathrm{exp}\left(-\dfrac{\eta_{1}^{2}}{2}\right),
\end{align}
with $\eta_{1}^{2}=4\pi\left(\frac{Z_{1}}{R_{Q}}\right)$ and

\begin{align}
\widetilde{\left\langle 0\right|}\mathrm{sin}\left(\varphi_{2}+\varphi_{x}^{(s)}\right)\widetilde{\left|1\right\rangle } & =\epsilon^{2}\left\langle 00\right|\mathrm{sin}\left(\varphi_{2}+\varphi_{x}^{(s)}\right)\left|01\right\rangle \nonumber\\
& =\epsilon^{2}\eta_{2}\mathrm{exp}\left(-\dfrac{\eta_{2}^{2}}{2}\right)\mathrm{cos}\left(\varphi_{\Delta}\right).\label{TransitionEqLast}
\end{align}

Using the WTQ example of Fig.\,\ref{GeneralWTQparams}(a)-(c), we compare in Figs.\,\ref{GeneralWTQparams}(d),(e) the
calculation for the relaxation and pure dephasing times as a function of normalized applied flux based on Eqs.\,(\ref{eq:T1}), (\ref{eq:Tphi}) and the diagonalization of the Hamiltonian
(Eq.\,(\ref{eq:2-deg-free-H})) versus the approximate results obtained using the analytical expressions for the matrix
elements derived above, where in both cases we assume a constant, real impedance $Z(\omega)=R$. To complete the picture, we plot in Fig.\,\ref{GeneralWTQparams}(f) the expected decoherence time $T_2$ for the same WTQ example given by the relation $T^{-1}_2=(2T_1)^{-1}+T^{-1}_{\phi}$. In our theoretical evaluation of the coherence times of the qubit, i.e., $T_1$, $T_{\phi}$, and $T_2$, we make the following assumptions, (1) the maximum relaxation time $100$ $\mu$s of the qubit is limited by a loss mechanism that is unrelated to the flux biasing source, such as dielectric loss. (2) the maximum decoherence time of the qubit $T_2$ is equal to $1.5T_1$ and limited by a non-flux noise source, such as thermal photon population in the readout resonator, (3) the parameters $R$ and $T$ associated with the flux biasing circuit are $0.01$ Ohm and $0.1$ K, respectively.   

Furthermore, to illustrate the dependence of the dephasing time on the parameter $R$ of the flux-biasing circuit for fixed $T=0.02$ K, we use the WTQ example of Fig.\,\ref{GeneralWTQparams} and plot in Fig.\,\ref{GeneralWTQdephasing}(a) the exact solution of $T_{\phi}$ versus normalized applied flux corresponding to varying parameter $R$. The blue, cyan, magenta, and red curves in Fig.\,\ref{GeneralWTQdephasing}(a) correspond to $R=1, 0.1, 0.05, 0.01$ Ohm, respectively. As expected, we find that the dips in $T_{\phi}$ away from the sweet spots increase with decreasing $R$. This is because the smaller the resistance in parallel with the current source is (see Fig.\,\ref{fig:WTQ-IB}), the larger the current portion flowing through it. 

Likewise, in Fig.\,\ref{GeneralWTQdephasing}(b), we illustrate the dependence of $T_{\phi}$ on the parameter $T$, while keeping $R=0.1$ Ohm constant. The blue, cyan, magenta, and red curves in Fig.\,\ref{GeneralWTQdephasing}(b) correspond to $T=0.02, 0.2, 0.4, 1$ K, respectively. As expected in this case as well, the dips in $T_{\phi}$ away from the sweet spots increase with $T$ of the flux-biasing circuit resistor that is in parallel with the current source (see Fig.\,\ref{fig:WTQ-IB}).     
 
\section{Experimental results}

\begin{figure*}
	[tb]
	\begin{center}
		\includegraphics[
		width=1.7\columnwidth 
		]%
		{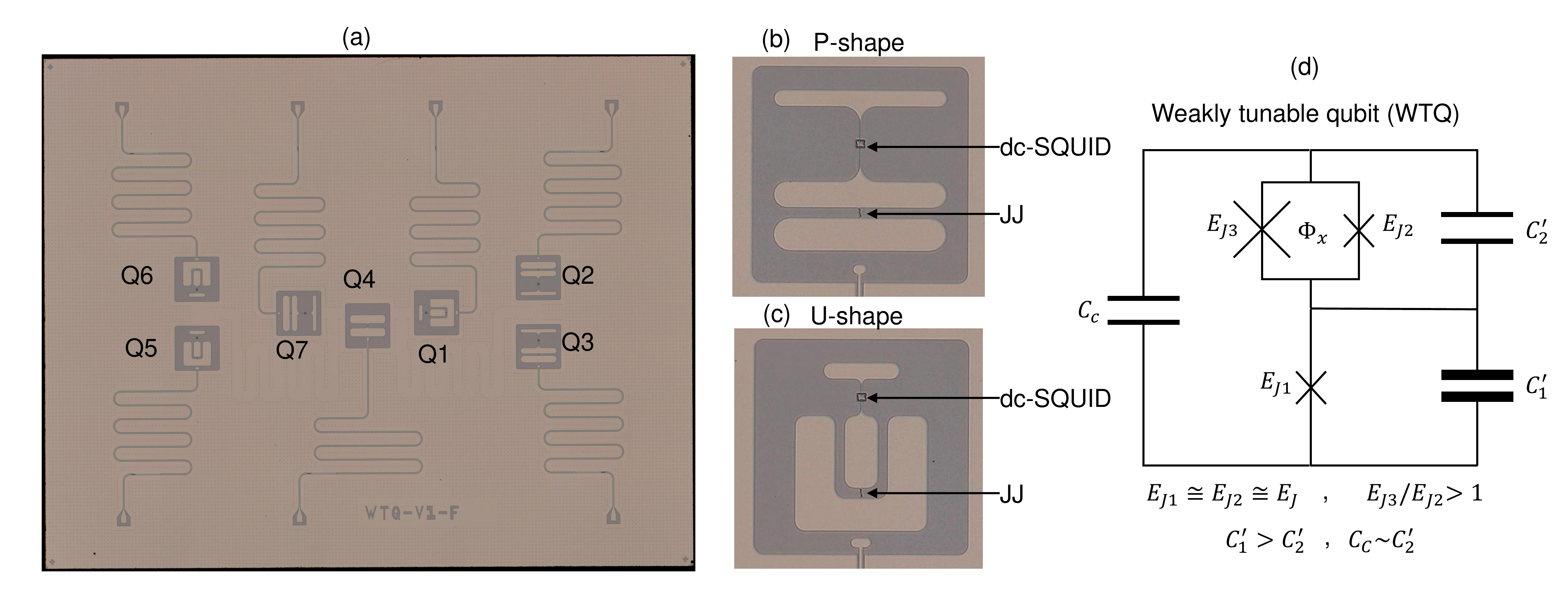}
		\caption{(a) A photo of one of the two 7-qubit chips measured in this work. The chip consists of 6 WTQs and one single-JJ transmon (Q4). The WTQs are realized using two gap-capacitance geometries shown in (b) and (c). (b) P-shape WTQ (Q2, Q3, Q7), in which the capacitance pads are parallel. (c) U-shape WTQ (Q1, Q5, Q6), in which one capacitance pad is curved. (d) Equivalent WTQ circuit. $C_1'$ and $C_2'$ represent the total capacitance shunting the JJs (including the self capacitance of the JJs). The small junction of the SQUID is comparable to the junction of the single-JJ transmon $E_{J2} \cong E_{J1} \cong E_{J}$, whereas the other JJ of the SQUID is slightly larger, giving $\alpha_J=2-5$. Rounded corners used in the electrodes are an effort to minimize E field concentration and are not specific to the WTQ design. 
		}
		\label{Chip}
	\end{center}
\end{figure*}

\begin{figure*}
	[tb]
	\begin{center}
		\includegraphics[
		width=1.7\columnwidth 
		]%
		{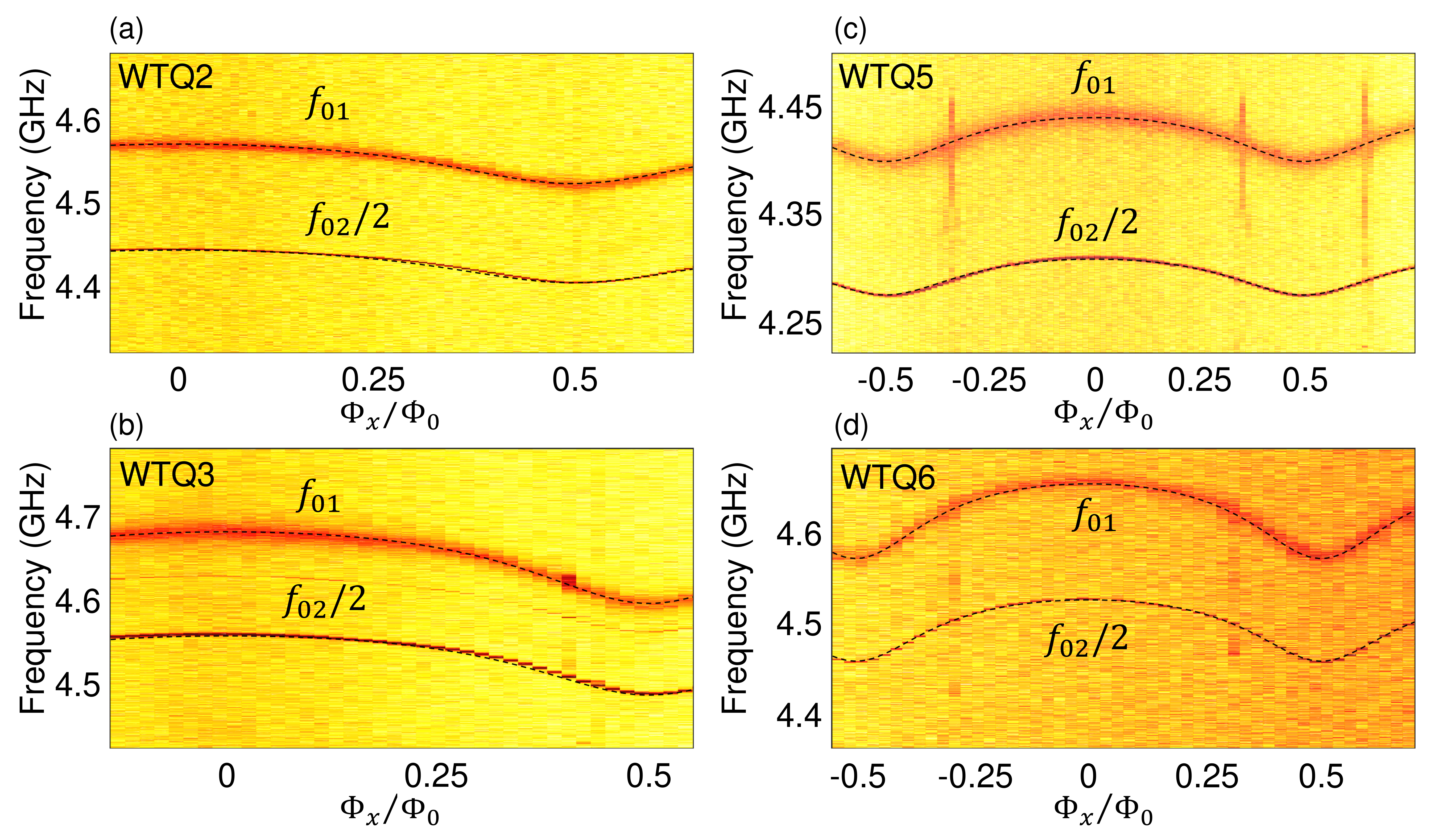}
		\caption{Representative qubit spectroscopy measurements of chip A plotted versus normalized external flux. The data shows $f_{01}$, and $f_{02}/2$ curves. (a) WTQ2 (P-shape). (b) WTQ3 (P-shape). (c) WTQ5 (U-shape). (d) WTQ6 (U-shape). WTQ2 and WTQ5, having $\alpha_J=3.5$, exhibit a small tunability of $50$ MHz and $43$ MHz, respectively. WTQ3 and WTQ6, having $\alpha_J=2.8$, exhibit a slightly larger tunability of $89$ MHz and $86$ MHz. Dashed black curves are theory fits.      
		}
		\label{SpectvsFlux}
	\end{center}
\end{figure*}

\begin{figure*}
	[tb]
	\begin{center}
		\includegraphics[
		width=2\columnwidth 
		]%
		{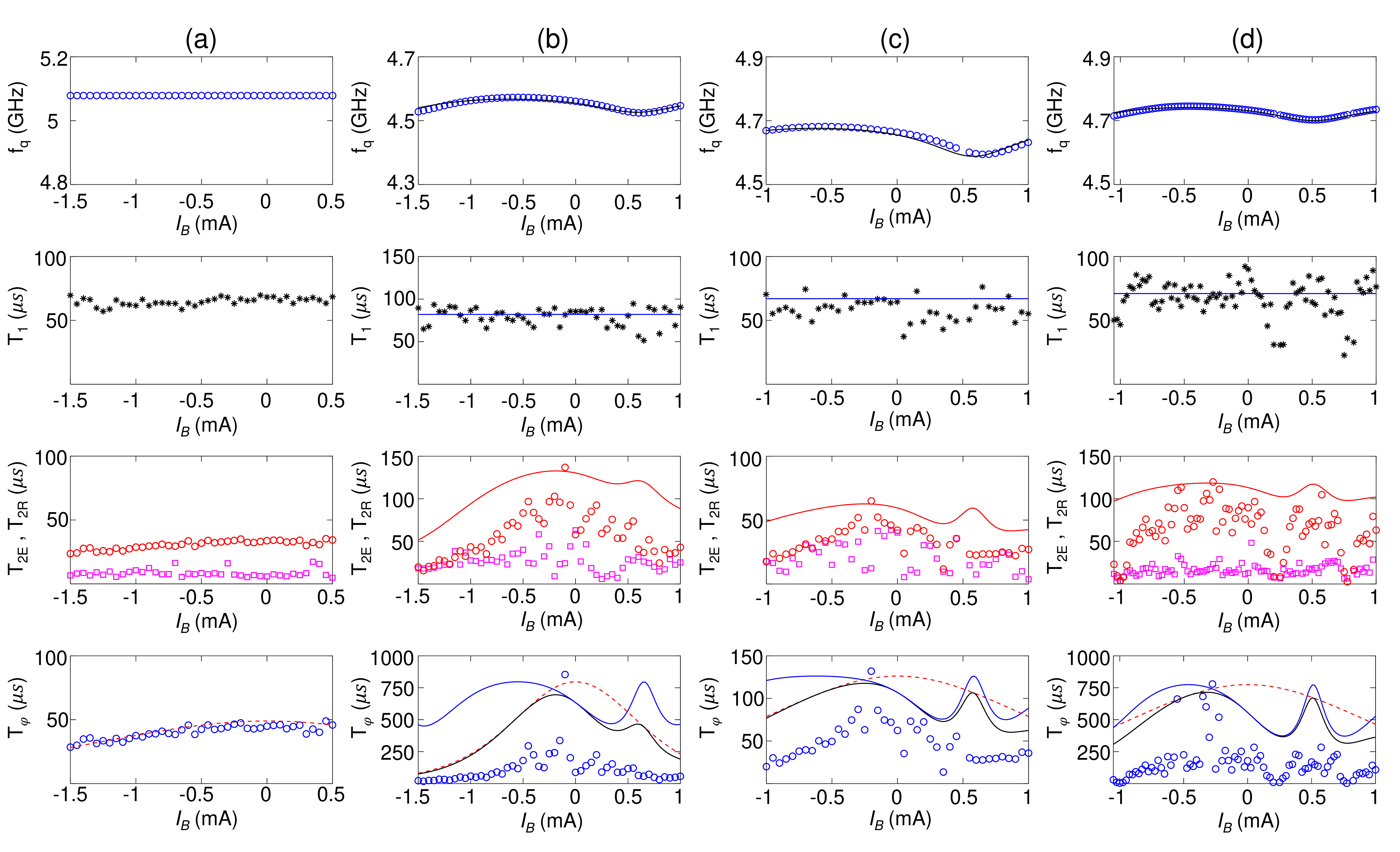}
		\caption{Coherence measurements for chip A taken for Q4 (a), WTQ2 (b), WTQ3 (c), and WTQ5 (d). Top to bottom rows show the qubit frequency $f_{\rm{q}}$ (blue circles), $T_1$ (black stars), $T_{\rm{2E}}$, $T_{\rm{2R}}$ (red circles and magenta squares, respectively), and $T_{\varphi}$ (blue circles) as a function of the applied coil bias current $I_{\rm{B}}$. The solid black curves in the first row plots represent theory fits for the qubit frequency based on the diagonalization of the Hamiltonian. The solid blue curves drawn in the second row plots, represent fits based on Eq.\,(\ref{eq:T1}). The solid red curves shown in the third row plots correspond to the calculated $T_{\rm{2E}}$ based on the $T_1$ and $T_{\varphi}$ fits (see main text). The dashed red, solid blue, and solid black curves shown in the fourth row plots represent the calculated $T_{\phi,\rm{D}}$, $T_{\varphi,\rm{F}}$, and $T_{\varphi}$ fits, respectively (see main text and table \,\ref{chip A dephasing} for details). }
		\label{ChipACoherence}
	\end{center}
\end{figure*}

\begin{figure*}
	[tb]
	\begin{center}
		\includegraphics[
		width=2\columnwidth 
		]%
		{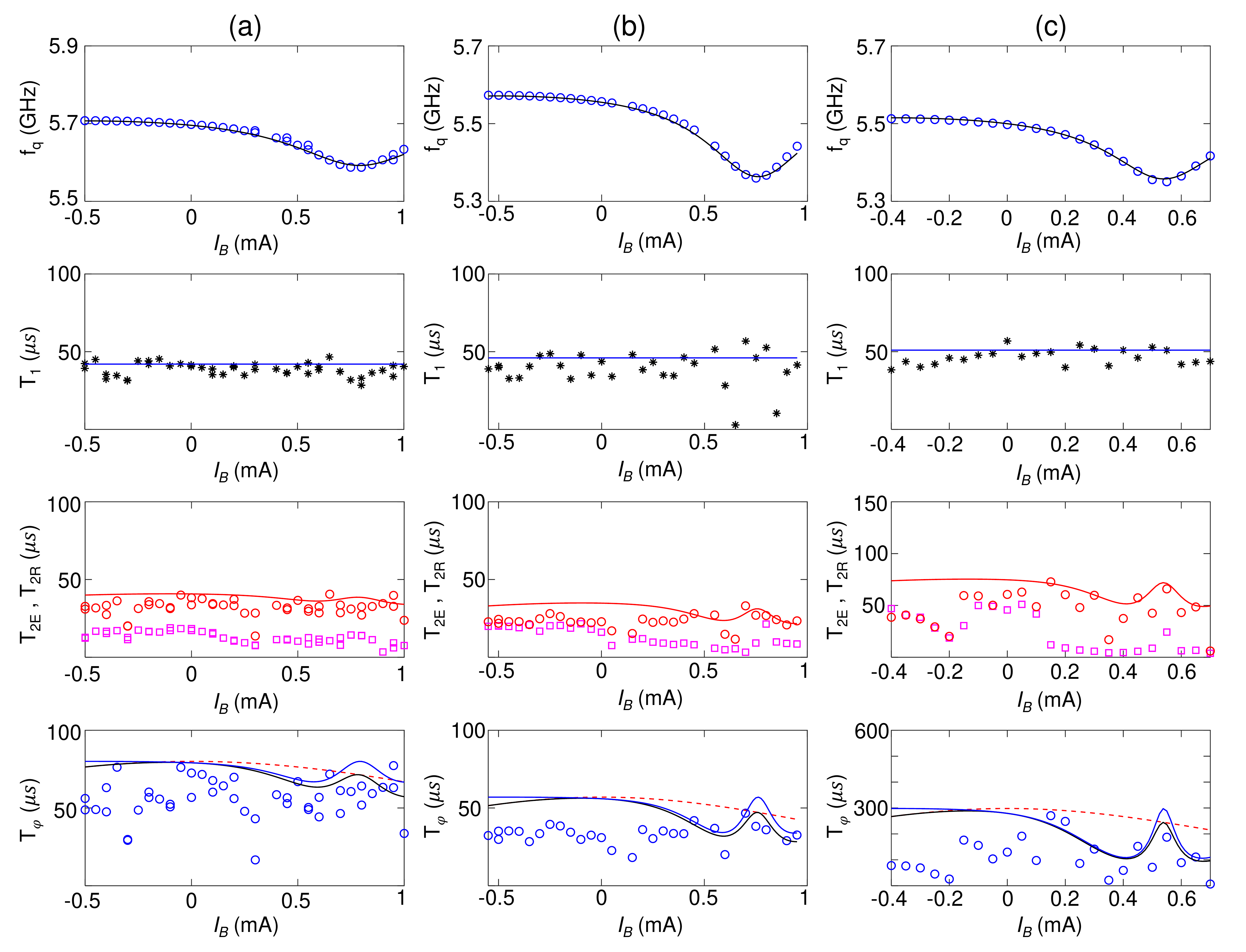}
		\caption{Coherence measurements for chip B taken for WTQ2 (a), WTQ3 (b), and WTQ5 (c). Top to bottom rows show the qubit frequency $f_{\rm{q}}$ (blue circles), $T_1$ (black stars), $T_{\rm{2E}}$, $T_{\rm{2R}}$ (red circles and magenta squares, respectively), and $T_{\varphi}$ (blue circles) as a function of the applied coil bias current $I_{\rm{B}}$. The solid black curves in the first row plots represent theory fits for the qubit frequency based on the diagonalization of the Hamiltonian. The solid blue curves drawn in the second row plots, represent fits based on Eq.\,(\ref{eq:T1}). The solid red curves shown in the third row plots correspond to the calculated $T_{\rm{2E}}$ based on the $T_1$ and $T_{\varphi}$ fits (see main text). The dashed red, solid blue, and solid black curves shown in the fourth row plots represent the calculated $T_{\phi,\rm{D}}$, $T_{\varphi,\rm{F}}$, and $T_{\varphi}$ fits, respectively (see main text and table \,\ref{chip B dephasing} for details).      
		}
		\label{ChipBCoherence}
	\end{center}
\end{figure*}

\begin{figure}
	[tb]
	\begin{center}
		\includegraphics[
		width=\columnwidth 
		]%
		{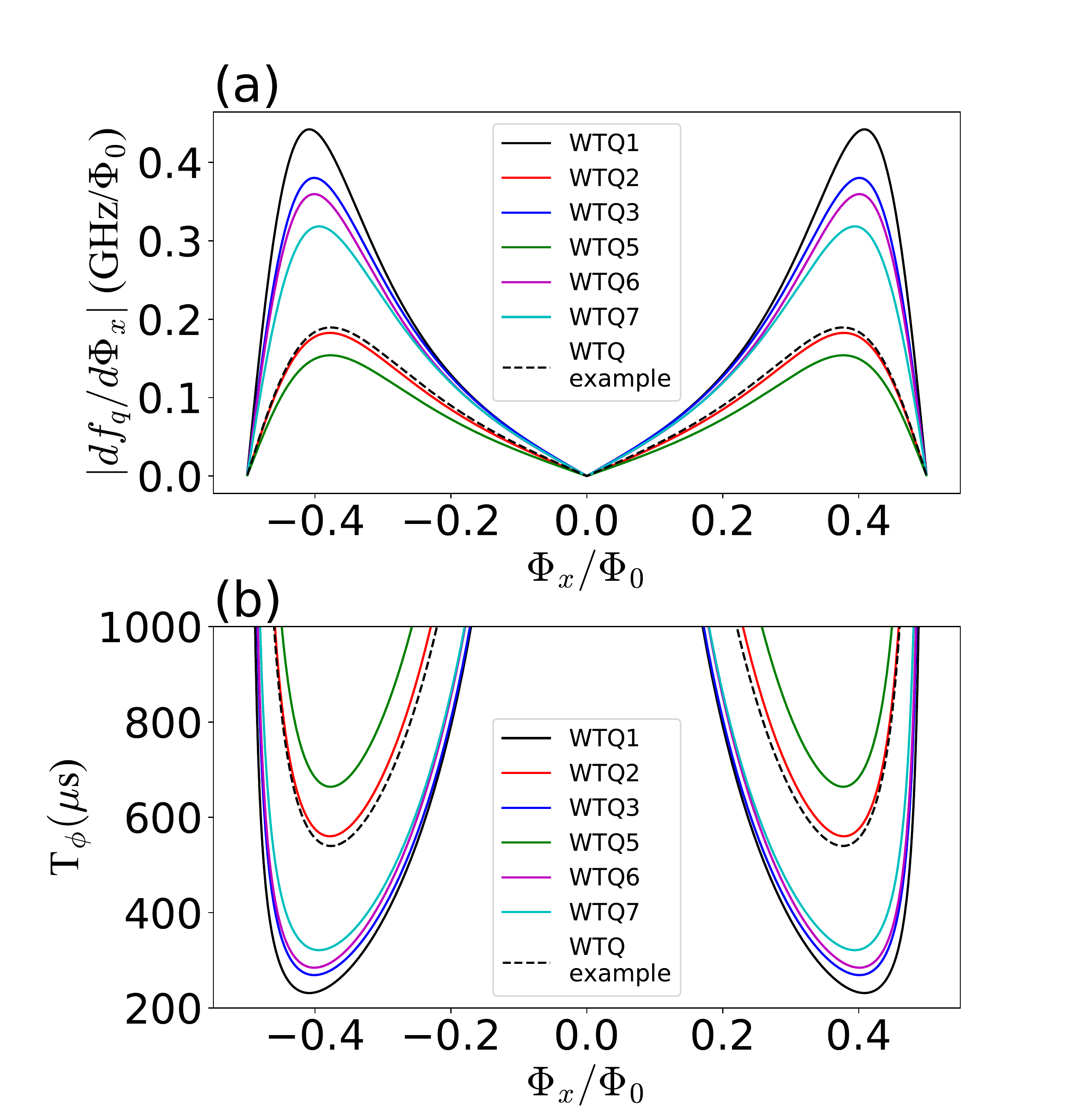}
		\caption{(a) WTQ flux-sensitivities of qubits on chip A and of example device, found by numerical differentiation of Eq.\,(\ref{eq:wq}), using device parameters listed in Table \,\ref{chip A param} and Fig.\,\ref{GeneralWTQparams}. (b) Flux-noise-limited dephasing times $T_{\phi}$, assuming $1/f$ flux-noise amplitude $A_{\Phi}^{1/2} \sim 2 \mu \Phi_0$ at $1$ Hz.      
		}
		\label{fig:slopesandTphis}
	\end{center}
\end{figure}

\begin{table*}[h!]
	\centering
	\begin{tabular}{c c c c c  c  c  c  c  c } 
		\hline
		Qubit & $\alpha_J$ & $f^{\rm{max}}_{01}$ (GHz) & $f_r$ (GHz) & $|\alpha|^{\rm{max}}$ (MHz) & $|\alpha|^{\rm{min}}$ (MHz) & $\delta$ (MHz) & $T_{\rm{1}}$ ($\mu$s) & $T^{\rm{max}}_{\rm{2E}}$ ($\mu$s) &  \\ [0.5ex] 
		\hline
		\textbf{WTQ1} & 2.6 & 4.8905 & 6.8869 & 254 & 224 & 99 & 75 & 85 \\ 
		\textbf{WTQ2} & 3.5 & 4.57 & 6.9642 & 254 & 233 & 50 & 82 & 136 \\
		\textbf{WTQ3} & 2.8 & 4.681 & 6.8372 & 246 & 215 & 89 & 64 & 65  \\
		\textbf{Q4} & NA & 5.0785 & 6.9567 & 360 & 360 & NA & 65 & 36 \\
		\textbf{WTQ5} & 3.5 & 4.442 & 6.9217 & 265 & 248 & 43 & 71 & 120  \\  
		\textbf{WTQ6} & 2.8 & 4.653 & 6.790 & 260 & 239 & 86 & - & -  \\ 
		\textbf{WTQ7} & 3.0 & 4.6065 & 6.867 & 252 & 226 & 76 & - & -  \\ 
		\hline 
	\end{tabular}
	\caption{Measured parameters of Chip A. Q4 is the single-JJ transmon on the chip. The SQUID JJ ratio parameter $\alpha_J$ is extracted from the spectroscopy fits. $f^{\rm{max}}_{01}$ is the maximum measured qubit frequency. $\alpha^{\rm{max}}$ ($\alpha^{\rm{min}}$) is the maximum (minimum) measured anharmonicity obtained at $f^{\rm{max}}_{01}$ ($f^{\rm{min}}_{01}$). $\delta$ is the measured frequency tunability range. $T^{\rm{max}}_{\rm{2E}}$ is the maximum measured coherence and $T_{1}$ is the corresponding lifetime taken at the same flux bias. The coherence times for WTQ6 and WTQ7 were not measured.}
	\label{chip A Meas}
\end{table*} 

\begin{table*}[t]
	\centering
	\begin{tabular}{c c c  c  c  c  c  c c c c} 
		\hline
		Qubit & Design & $\alpha_S$  & $C_1'$ (fF) & $C_2'$ (fF) & $C_c$ (fF) & $I_{\rm{c1}}$ (nA) & $I_{\rm{c2}}$ (nA) & $I_{\rm{c3}}$ (nA) & $\alpha_J$ (fit) & \\ [0.5ex] 
		\hline
		\textbf{WTQ1} & U & 4.2 & 60.5 & 17.8 & 20.0 & 28.3 & 25.0 & 65.3 & 2.6 \\ 
		\textbf{WTQ2} & P & 5.1 & 61 & 18.4 & 20.5 & 25.2 & 20.0 & 69.8 & 3.5 \\
		\textbf{WTQ3} & P & 4.2 & 61 & 17.8 & 20.6 & 26.4 & 21.3 & 60 & 2.8  \\
		\textbf{Q4} & NA & NA & 62.9 & NA & NA & 24.0 & NA & NA & NA \\
		\textbf{WTQ5} & U & 5.1 & 60.0 & 18.4 & 20.0 & 23.4 & 20.2 & 70.7 & 3.5 \\  
		\textbf{WTQ6} & U & 4.2 & 60.7 & 18.3 & 20.1 & 25.9 & 21.3 & 60.0 & 2.8 \\ 
		\textbf{WTQ7} & P & 4.2 & 60.0 & 18.1 & 20.7 & 25.4 & 20.0 & 60.2 & 3.0 \\ 
		\hline 
	\end{tabular}
	\caption{Circuit parameters of chip A qubits extracted from the spectroscopy theory fits.}
	\label{chip A param}
\end{table*}

\begin{table*}[h!]
	\centering
	\begin{tabular}{c c c  c c c c c  c  c } 
		\hline
		Qubit & $\alpha_J$ & $f^{\rm{max}}_{01}$ (GHz) & $f_r$ (GHz) & $|\alpha|^{\rm{max}}$ (MHz) & $|\alpha|^{\rm{min}}$ (MHz) & $\delta$ (MHz) & $T_{\rm{1}}$ ($\mu$s) & $T^{\rm{max}}_{\rm{2E}}$ ($\mu$s) &  \\ [0.5ex] 
		\hline
		\textbf{WTQ2} & 2.36 & 5.6805 & 6.9683 & 252 & 224 & 115 & 42 & 40 \\
		\textbf{WTQ3} & 2.06 & 5.557 & 6.8410 & 243 & 189 & 207 & 57 & 33  \\
		\textbf{Q4} & NA & 6.375 & 6.967 & 349 & 349 & NA & 13 & 25 \\
		\textbf{WTQ5} & 2.06 & 5.497 & 6.9262 & 250 & 161 & 159 & 47 & 73  \\  
		\textbf{WTQ6} & 1.9 & 5.743 & 6.7906 & 248 & 181 & 287 & 40 & 58  \\ 
		\textbf{WTQ7} & 1.95 & 5.972 & 6.869 & 226 & 162 & 262 & 29 & 42  \\ 
		\hline 
	\end{tabular}
	\caption{Measured parameters of Chip B. Q4 is a single-JJ transmon. WTQ1 did not yield (one of its SQUID junctions is open). The SQUID JJ ratio parameter $\alpha_J$ is extracted from the spectroscopy fits. $f^{\rm{max}}_{01}$ is the maximum measured qubit frequency. $\alpha^{\rm{max}}$ ($\alpha^{\rm{min}}$) is the maximum (minimum) measured anharmonicity obtained at $f^{\rm{max}}_{01}$ ($f^{\rm{min}}_{01}$). $\delta$ is the measured frequency tunability range. $T^{\rm{max}}_{\rm{2E}}$ is the maximum measured coherence and $T_{1}$ is the corresponding lifetime taken at the same flux bias. The relatively low $T_1$ of Q4 is limited by the Purcell effect due to the proximity of $f_{01}$ to the readout frequency.}
	\label{chip B Meas}
\end{table*}

\begin{table*}[h!]
	\centering
	\begin{tabular}{c c c  c  c  c  c  c c c c} 
		\hline
		Qubit & Design & $\alpha_S$ & $C_1'$ (fF) & $C_2'$ (fF) & $C_c$ (fF) & $I_{\rm{c1}}$ (nA) & $I_{\rm{c2}}$ (nA) & $I_{\rm{c3}}$ (nA) & $\alpha_J$ (fit) & \\ [0.5ex] 
		\hline
		\textbf{WTQ2} & P & 3 & 61.4 & 18.3 & 20.0 & 37.9 & 40.0 & 94.4 & 2.36 \\
		\textbf{WTQ3} & P & 2.9 & 61.5 & 18.5 & 20.7 & 36.9 & 36.0 & 74 & 2.06  \\
		\textbf{Q4} & NA & NA & 62.9 & NA & NA & 36.7 & NA & NA & NA \\
		\textbf{WTQ5} & U & 3 & 61.4 & 18.0 & 20.0 & 35.7 & 40.0 & 82.4 & 2.06 \\  
		\textbf{WTQ6} & U & 2.9 & 60.3 & 18.3 & 20.0 & 38.5 & 38.0 & 72.0 & 1.9 \\ 
		\textbf{WTQ7} & P & 2.9 & 60.1 & 18.0 & 20.7 & 42.0 & 41.0 & 80.0 & 1.95 \\ 
		\hline 
	\end{tabular}
	\caption{Circuit parameters of chip B qubits extracted from the spectroscopy theory fits. WTQ1 on chip B is not measured because one of its SQUID junctions did not yield (open).}
	\label{chip B param}
\end{table*}

\begin{table*}[h!]
	\centering
	\begin{tabular}{c c c  c  c  c  c } 
		\hline
		Qubit & $\kappa/2\pi$ (MHz) & $\chi/2\pi$ (MHz) & $\bar{T}_e$ (mK) & $\varTheta$ $\rm{mK}/(\rm{mA})^2$ & $\bar{T}_m$ (K) & $R$ (m$\Omega$)  \\ [0.5ex] 
		\hline
		\textbf{WTQ2} & 0.45 & 0.47 & 47 & 10 & 0.2 & 1  \\
		\textbf{WTQ3} & 0.65 & 0.39 & 67 & 7 & 0.2 & 1   \\
		\textbf{Q4} & 0.82 & 0.51 & 78 & 5 & 0.2 & 1  \\
		\textbf{WTQ5} & 0.65 & 0.26 & 55 & 5 & 0.2 & 1\\  
		\hline 
	\end{tabular}
	\caption{Parameters of chip A qubits used in the coherence measurement theory fits shown in Fig.\,\ref{ChipACoherence}}
	\label{chip A dephasing}
\end{table*}

\begin{table*}[tbh]
	\centering
	\begin{tabular}{c c c  c  c  c  c } 
		\hline
		Qubit & $\kappa/2\pi$ (MHz) & $\chi/2\pi$ (MHz) & $\bar{T}_e$ (mK) & $\varTheta$ $\rm{mK}/(\rm{mA})^2$ & $\bar{T}_m$ (K) & $R$ (m$\Omega$)  \\ [0.5ex] 
		\hline
		\textbf{WTQ2} & 0.6 & 0.4 & 74 & 3 & 0.1 & 2  \\
		\textbf{WTQ3} & 0.5 & 0.63 & 70 & 5 & 0.1 & 2   \\
		\textbf{WTQ5} & 0.65 & 0.39 & 58 & 7 & 0.05 & 2\\  
		\hline 
	\end{tabular}
	\caption{Parameters of chip B qubits used in the coherence measurement theory fits shown in Fig.\,\ref{ChipBCoherence}}
	\label{chip B dephasing}
\end{table*}

We realize and measure two seven-qubit chips (referred to as A and B), which are similar in design to those with single JJ-transmons we measured in the past \cite{MultiqubitMaika}. Each chip includes six WTQs and one single-JJ transmon as shown in the device photo in Fig.\,\ref{Chip}(a). Each qubit is capacitively coupled to a readout resonator, which, in turn, is  capacitively coupled to a readout port. All resonator buses coupling the qubits are disabled by shortening their ends to ground. We implement the WTQs in two configurations, dubbed P-shape and U-shape, which differ in the shape of the gap-capacitance electrodes shunting the JJs. In the P-shape configuration exhibited in Fig.\,\ref{Chip}(b), the three capacitance electrodes are parallel to each other, whereas in the U-shape configuration displayed in Fig.\,\ref{Chip}(c), one of the outer electrodes wraps around three sides of the middle electrode of the qubit. The motivation for designing WTQs using these two possible configurations is to experimentally examine if either holds any coherence advantage over the other. In Fig.\,\ref{Chip}(d), we exhibit a simplified circuit of the WTQ. The shunting capacitances $C_1'$, $C_2'$ include the capacitances of the JJs. Figure\,\ref{Chip}d also outlines the useful parameter space of WTQ capacitances and Josephson energies (i.e., in comparison to the transmon qubits presented in Fig.\,\ref{Transmons}).

In what follows, we present the main results measured for chip A and B, whose circuit parameters are listed in Tables \,\ref{chip A Meas}, \ref{chip A param} for chip A and Tables \,\ref{chip B Meas}, \ref{chip B param} for chip B. The WTQs in chip A yield a smaller $\delta$ range $43-99$ MHz (see Table \,\ref{chip A Meas}) versus $115-288$ MHz (see Table \,\ref{chip B Meas}) for chip B, since they are designed with a slightly higher JJ area ratio of the SQUIDs, i.e., $\alpha_S=A_{J_3}/A_{J_2}$, where $A_{J_2}$ and $A_{J_3}$ represent the design area of the respective junctions. In chip A, we set $\alpha_S=4.2$ for WTQ1, WTQ3, WTQ6, WTQ7, versus $\alpha_S=2.9$ for the corresponding qubits in chip B, and $\alpha_S=5.1$ for WTQ2 and WTQ5 in chip A versus $\alpha_S=3.0$ in chip B. The two chips also differ in the JJ oxidation conditions applied in fabrication, yielding higher maximum WTQ frequencies for chip B, i.e., $5.5-5.97$ GHz, than chip A, i.e., $4.44-4.89$ GHz. Lastly, chips A and B are measured in two different cooldowns using different dc wiring between the $4$ K stage and the base-temperature stage ($20$ mK) of the dilution fridge. In the case of chip A, the intermediate dc wiring is superconducting while in the case of chip B, it is resistive.    

In Fig.\,\ref{SpectvsFlux}(a)-(d), we show spectroscopy measurements of WTQ2 (P-shape), WTQ3 (P-shape), WTQ5 (U-shape), and WTQ6 (U-shape) of chip A, respectively, plotted versus the normalized external flux threading the SQUID loop. The magnetic flux is generated using a dc-current $I_B$ applied to a small global superconducting coil with $L_c\backsimeq5.5$ mH attached to the device copper package. The mutual inductance $M$ between the coil and the SQUID loops across the chip vary in the range $0.8-1.1$ pH. The dashed black curves plotted over the data represent theoretical fits for the qubit frequency $f_{01}$ and $f_{02}/2$. In Table \,\ref{chip A param}, we list the capacitance and critical current parameters of the various qubits used in the spectroscopy fits  (including those of WTQ1 and WTQ7 whose data are not shown). The fits also allow us to extract the JJ asymmetry parameter of the dc-SQUIDs $\alpha_J$ for the various qubits, which we list in Tables \,\ref{chip A Meas}, \ref{chip A param} for chip A. As expected from the device physics, higher values of $\alpha_J$ correlate well with smaller observed $\delta$ of the WTQs.   

As seen in the figure, the maximum and minimum qubit frequencies are obtained at $\Phi_{x}=0$ and $\Phi_{x}=\pm \Phi_0/2$, respectively. The smallest $\delta$ of $50$ MHz and $43$ MHz are measured for WTQ2 and WTQ5, respectively, which are designed to have a slightly larger JJ asymmetry than WTQ3 and WTQ6, whose $\delta$ is $89$ MHz and $86$ MHz, respectively. Notably, the $\alpha_J$ extracted from the fits (see Table \ref{chip A param}), i.e., $3.5$ and $2.8$, match fairly well the dc resistance ratios measured for test JJs (fabricated on the same wafer) that have the same area ratios as the JJs of the SQUIDs. 

Furthermore, using the spectroscopy data, we calculate the maximum and minimum magnitude of the WTQ anharmonicity given by $|\alpha|=|f_{12}-f_{01}|$ at the upper and bottom sweet spots corresponding to $\Phi_{x}=0$ and $\Phi_{x}=\pm \Phi_0/2$, respectively (see Table \,\ref{chip A Meas}). For example, the anharmonicity of qubits with $\delta$ of $43-99$ MHz varies by $17-31$ MHz.

Similarly, Table \,\ref{chip B Meas} summarizes the main figures of merit measured for qubits in chip B, while Table \,\ref{chip B param} lists the capacitance and critical current parameters of the various qubits in chip B extracted from their corresponding spectroscopy fits (not shown). 

In Figs.\,\ref{ChipACoherence}, \,\ref{ChipBCoherence}, we plot the measured coherence times for qubits in chip A and chip B respectively. In Fig.\,\ref{ChipACoherence}, the columns, from left to right, represent the results of the single-JJ transmon Q4 (a), WTQ2 (b), WTQ3 (c), and WTQ5 (d). The rows, from top to bottom, display the qubit frequency $f_{\rm{q}}$, relaxation time $T_1$, decoherence times $T_{\rm{2E}}$ (Echo) and $T_{\rm{2R}}$ (Ramsey), and extracted dephasing time given by $T_{\rm{\varphi}}^{-1}=T_{\rm{2E}}^{-1}-(2T_{\rm{1}})^{-1}$, plotted as a function of $I_B$, i.e., the direct current applied to the global coil flux biasing the qubits. 

The solid black curves in the first row plots represent theory fits based on the diagonalization of the WTQ Hamiltonian. The solid blue curves in the second row plots represent theory fits for $T_1$ calculated using Eq.\,(\ref{eq:T1}). The upper bound of these fits is set to match the experimental data (typically $T_1$ that corresponds to the maximum $T_{\rm{2E}}$, listed in Table \,\ref{chip A Meas} (for chip A and Fig.\,\ref{ChipACoherence}) and Table \,\ref{chip B Meas} (for chip B and Fig.\,\ref{ChipBCoherence})). The average $T_1$ of the WTQs is similar to that of the fixed-frequency transmon Q4. This is consistent with the assumption that surface dielectric loss dominates relaxation in both the WTQ and the single-JJ transmon. Variations with flux as seen for instance in chip A, WTQ5 (Fig.\,\ref{ChipACoherence})) also suggest frequency-dependent couplings to TLSes. However we did not explore this behavior systematically as was done in Ref. \cite{DynamicsT1}.

One important observation regarding the measured $T_{\rm{2E}}$ (red circles) and $T_{\rm{2R}}$ (magenta squares) plotted in the third row and $T_{\rm{\varphi}}$ (blue circles) drawn in the fourth row of Fig.\,\ref{ChipACoherence}, is that they exhibit a pronounced decrease with $|I_B|$, which strongly indicates that the decoherence times in our system are primarily limited by heating effects caused by large $|I_B|$. This observation is further supported by the fact that  we observe a rise in the temperature of the mixing chamber stage by several milliKelvin during the lengthy data taking process of $f_q$, $T_1$, $T_{\rm{2R}}$, $T_{\rm{2E}}$ as $I_B$ is swept (about $15$ minutes for each $I_B$). Another supporting evidence of the heating effect we observe in our system, is the monotonic decrease in $T_{\rm{2R}}$, $T_{\rm{2E}}$, and  $T_{\rm{\varphi}}$ of the single JJ-transmon (Q4) (see the first column plots), which to a large extent is insensitive to flux noise.  

To model the dependence of the dephasing time on heating caused by $|I_B|$, we consider one of the dominant dephasing mechanisms in qubits that are dispersively coupled to readout resonators, which arise from fluctuation in the qubit frequency due to thermal photon population in the readout resonator \cite{ClerkShotNoise}. In the limit $\bar{n}\ll1$, the dephasing rate associated with this mechanism (denoted $\Gamma_{\rm{\phi,D}}\equiv T_{\rm{\phi,D}}^{-1}$) is given by \cite{CavityAtten}
        
\begin{equation}
\Gamma_{\rm{\phi,D}}=\Gamma_c\bar{n},\label{DephasingRate2}\\
\end{equation}  

\noindent where $\Gamma_c\equiv\kappa\chi^2/(\kappa^2+\chi^2)$, $\kappa$ is the total photon decay rate of the fundamental mode of the readout resonator with angular frequency $\omega_r=2\pi f_r$ (here $\kappa$ is dominated by the coupling rate to the external feedline), $\chi$ is the qubit-state-dependent frequency shift of the readout resonator, and $\bar{n}$ is the average thermal photon number in the resonator, where  $\bar{n}=1/\left(e^{\left(\hbar\omega_r/k_BT_e\right)}-1\right)$ is the Bose-Einstein population of the $50$ Ohm external feedline (heat bath) at effective temperature $T_e$, which we express as $T_e=\bar{T}_e+\delta T$, where $\bar{T}_e$ is the effective temperature with no bias current ($I_B=0$) and $\delta T=\Theta I_B^2$ represents the rise in the effective temperature of the device due to ohmic dissipation. 

Note that $\delta T$ can be expressed as $\delta T=C_HQ_d$, where $C_H$ is the heat capacity of the device and $Q_d=RI_B^2\tau_B/2$ is the dissipated heat energy in the flux biasing circuit, the experimentally relevant temperature-to-current conversion coefficient $\Theta$ can be expressed as $\Theta=RC_H\tau_B/2$, where $\tau_B$ is an effective measurement duration. In our qubit experiments, we find that $\Theta$ varies in the range $5-10$ $\rm{mK}/{\rm{(mA)}}^2$ for chip A (see Table \,\ref{chip A dephasing}) and $3-7$ $\rm{mK}/{\rm{(mA)}}^2$ for chip B (see Table \,\ref{chip B dephasing}).         

The dashed red curves in the fourth row plots of Fig.\,\ref{ChipACoherence} represent the calculated $T_{\rm{\phi,D}}$ for Q4 (a), WTQ2 (b), WTQ3 (c), and WTQ5 (d). Similarly, the solid blue curves represent the calculated dephasing time $T_{\rm{\varphi,F}}$ that is set by the flux noise of the biasing circuit given by  $T_{\rm{\varphi,F}}^{-1}=T_{\rm{\phi}}^{-1}+T_{\rm{\phi,D_0}}^{-1}$, where $T_{\rm{\phi}}$ is evaluated using Eq.\,(\ref{eq:Tphi}) with $T=\bar{T}_m+\delta T$, where $\bar{T}_m$ is the effective temperature of the superconducting magnetic coil with no bias current, and $T_{\rm{\phi,D_0}}$ is an experimental bound on dephasing time due to the dispersive coupling mechanism. It is worth noting that heating has a lesser effect on $T_{\rm{\varphi,F}}$ (solid blue curves) than $T_{\rm{\phi,D}}$ (dashed red curves). In the former case, heating causes the periodic response of $T_{\rm{\phi}}$ to decrease with $|I_B|$. Likewise, the solid black curves, drawn in the forth row plots, represent the total dephasing time calculated using the relation $T_{\rm{\varphi}}^{-1}=T_{\rm{\phi}}^{-1}+T_{\rm{\phi,D}}^{-1}$, which accounts for the contribution of both dephasing mechanisms discussed above and exhibits a fair agreement with the data. 

Finally, we plot a bound on $T_{\rm{2E}}$, drawn as solid red curves in the third row plots of Fig.\,\ref{ChipACoherence},  using the calculated $T_{\rm{\varphi}}$ and the relation $T_{\rm{2E}}^{-1}=(2T_1)^{-1}+T_{\rm{\varphi}}^{-1}$.   

In a similar manner, in Fig.\,\ref{ChipBCoherence} we plot using the same symbols and conventions of Fig.\,\ref{ChipACoherence} tunability curves and coherence time measurements taken for qubits WTQ2 (a), WTQ3 (b), and WTQ5 (c) of chip B as a function of $I_B$ and the corresponding theoretical fits.

\section{Discussion}

As seen in Figs.\,\ref{ChipACoherence}, \ref{ChipBCoherence}, the WTQs exhibit coherence times, i.e., $T_1$ and $T_{\varphi}$, that are comparable to those of single-JJ transmons fabricated on the same chip. These times are consistent with the loss and dephasing mechanisms typically seen in single-junction transmons, i.e., surface dielectric loss \cite{TransmonSurfaceLoss} and dispersive coupling in the case of $T_{\varphi}$ near the sweet spots (see the solid blue curves in the fourth row of Figs.\,\ref{ChipACoherence}, \ref{ChipBCoherence}). 

We attribute the larger than expected drop in the measured $T_{\rm{\varphi}}$ of the WTQs and transmons, observed for large $|I_B|$ (see blue circles), to the  unintended heating side-effect produced by our flux biasing circuit as $|I_B|$ is varied. This undesired heating effect is extrinsic to the qubits and can in principle be mitigated by better thermal anchoring of the dc wires and the magnetic coil, reducing the parasitic resistance of the superconducting coils and the intermediate connectors carrying the dc current at the base-temperature stage, and realizing on-chip, superconducting flux lines with higher mutual inductance to the SQUID loops that directly flux-bias each qubit, thus requiring much smaller bias currents. 

It is important to emphasize that the resistance of the flux biasing circuit affects the two dephasing mechanisms associated with our devices differently as reflected in the theoretical results of Fig.\,\ref{GeneralWTQdephasing}(a) and the experimental data in Figs.\,\ref{ChipACoherence}, \ref{ChipBCoherence}. As seen in Fig.\,\ref{GeneralWTQdephasing}(a) a higher resistance dampens the swings of the dephasing times at and away from the flux sweet spots, whereas higher resistance would increase the heating effect as seen in the experiment for large $|I_B|$. However, by applying some of the techniques outlined above, it should be possible to resolve this trade-off by reducing the heating effect in future systems. 

While we do not observe flux-noise-limited dephasing in our experiment, we can estimate its effects. So-called `universal' flux noise with an approximately $1/f$ power spectrum has been widely observed in superconducting systems. Its amplitude $A_{\Phi}^{1/2}$ at $1$ Hz generally exceeds $1$ $\mu \Phi_0$. \cite{wellstood_lowfrequency_1987, anton_pure_2012}. Following \cite{AsymmonsExp}, we can write an approximate Ramsey dephasing rate $\Gamma_{\phi,R} = |df_q / d \Phi_x| \cdot 2 \pi \sqrt{A_{\Phi} | \ln (2 \pi f_\textrm{IR} t) |}$, where $f_\textrm{IR}$ is a cutoff frequency of $1$ Hz, and $t$ being on the order of $1/\Gamma_{\phi,R}$ we take to be $10$ $\mu$s. We can find the flux sensitivity $|df_q / d \Phi_x|$ as a function of flux by differentiating Eq.\,(\ref{eq:wq}) with respect to flux. For example, for the qubits on chip A, we can use circuit parameters in Table \,\ref{chip A param}  and differentiate numerically in terms of fractional flux quantum, with units of GHz/$\Phi_0$, as shown in figure \ref{fig:slopesandTphis} (a). If we conservatively assume $A_{\Phi}^{1/2} \sim 2 \mu \Phi_0$, and taking the echoed dephasing rate $\Gamma_{\phi,E}$ according to \cite{bylander_noise_2011} to be $\Gamma_{\phi,E} \sim \frac{1}{4} \Gamma_{\phi,R}$, and neglecting the contribution of qubit relaxation, we can estimate a lower bound for flux-noise-limited dephasing time $T_\phi \sim 1/\Gamma_{\phi,E}$. These dephasing times, shown in figure \ref{fig:slopesandTphis} (b) compare favorably to $T_2^e = 200$ $\mu$s which is projected to be the minimum necessary to achieve fidelity $\geq 99.9\%$ in cross-resonance gates \cite{ku_suppression_2020}. The minimum dephasing times are comparable to the best average transmon dephasing times demonstrated in multi-qubit devices \cite{berg_probabilistic_2022}. Using the example parameters of Fig.\,\ref{GeneralWTQparams} we find a minimum flux-noise-limited dephasing time comparable to the dephasing times presented in Figs.\,\ref{GeneralWTQparams} and \ref{GeneralWTQdephasing}.

It is worth noting that we do not observe an obvious advantage for either the P or U WTQ designs with respect to the measured relaxation time $T_1$ of the various qubits in both chips (or at least not in the range $50-100$ $\mu$s measured in the experiment). 

It is also worth noting that, like single-JJ transmons, it is possible to adjust the maximum frequency of WTQs using focused laser beams following fabrication and JJ resistance measurement at room temperature \cite{laserAnnealing}. This useful technique is applicable to WTQs because (1) the maximum frequency of WTQs is strongly dependent on the critical current of the non-SQUID JJ, (2) the non-SQUID JJ of WTQs is physically well separated from the SQUID in both designs, thus it may be annealed without considerably affecting the SQUID JJs. Such capability further enhances the ability of WTQs to avoid frequency collisions in large quantum systems as laser annealing can be selectively applied to WTQs across the chip (whose resistances are far off from their designed values), followed by, as necessary, an \textit{in situ} fine-tuning using applied flux when the device is cold. Hence, selective laser-anneal, combined with in-situ flux tuning of WTQs, suggests a new avenue to reliably evade frequency collisions in large lattices of qubits, while maintaining high coherence. To quantify the effect on frequency-crowding, we can adopt the simplified model \cite{laserAnnealing} that a `collision free' device of $N$ qubits requires every qubit to lie within a `window' $\pm \Delta f$ of its frequency set-point. For normally distributed scatter at frequency precision $\sigma_f$, the likelihood of this occurring is found as the cumulative distribution function of $\Delta f / \sigma_f$, raised to the power $N$. For the conditions described in \cite{laserAnnealing}, a heavy-hexagon-type lattice of $N = 1000$ qubits has $\Delta f = 26$ MHz. WTQ tunability $\delta$ however in practice enlarges this `window' by $\delta/2$. Taking the precision $\sigma_f = 18$ MHz shown for multi-qubit lattices in \cite{laserAnnealing2}, we estimate that a $1000$-qubit lattice made of WTQs of tunability $\delta = 50$ MHz can be made collision-free 10\% of the time, while using WTQs of tunability $\delta = 99$ MHz gives a collision-free yield of $99$\%.
 
Following this work, there are several promising threads to pursue, such as: (1) Investigating the dependence of higher modes of the WTQ on applied flux, in particular the $f_{10}$ mode that is theoretically calculated for the WTQ example of Fig.\,\ref{GeneralWTQparams}(c). (2) Demonstrating cross-resonance gates in multi-qubit chips of WTQs and showing that most frequency collisions can be avoided by applying independent fluxes to neighboring qubits. (3) Finding out the limits on the tunability ranges that can be experimentally achieved with these qubits. (4) Interrogating TLS spectra in the frequency and time domains by flux-tuning a WTQ.    

\section{Conclusion}

We introduce weakly tunable superconducting qubits whose frequency can be tuned with external magnetic flux. The qubits comprise capacitively shunted JJ and asymmetric dc-SQUID, sharing one electrode and capacitively coupled via the other two. 

We develop a theoretical model that captures the device physics and its coupling to the flux biasing circuit. By solving the full Hamilitonian of the system, we calculate the various qubit properties as a function of the circuit parameters and applied flux. We also calculate the qubit relaxation and dephasing times associated with the flux-biasing circuit. Furthermore, we derive analytical expressions that yield approximate values for the various device parameters and coherence times. 

Moreover, we fabricate and test two superconducting chips containing several variations of these qubits. We show that they can achieve frequency tunablity ranges as low as $43-287$ MHz with only a small asymmetry in the size of their SQUID JJs ($\alpha_J=2-3.5$), which are considerably lower than what is possible using highly-asymmetric dc-SQUID transmons (about $350$ MHz of tunability with an asymmetry factor of $15$). Using such weakly flux-tunable qubits should enable us to resolve most common frequency-collisions in multi-qubit architectures while minimizing sensitivity to flux noise. For example, a $1000$-qubit device comprising WTQs of tunability $\delta = 50$ MHz and trimmed using laser-anneal could achieve collision-free yield of $10$\% and dephasing times $> 500$ $\mu$s, while WTQs of tunability $\delta = 99$ MHz could achieve collision-free yield of $99$\% and dephasing times $> 200$ $\mu$s. Furthermore, in any large qubit lattice it is likely that several qubits will suffer degraded coherence due to TLS coupling. WTQ tunability can restore the qubits' $T_1$ and permit periodic adjustments if the TLSes drift over time \cite{DynamicsT1}.  

These qubits also retain key properties of transmons, that make them quite useful in multi-qubit architectures, such as having $0-1$ transition frequencies in the range $4-5.5$ GHz, anharmonicities around $250-300$ MHz, comparable relaxation and decoherence times to single-JJ transmons, small footprints, and a standard fabrication process. Another important advantage of these qubits is that they couple together like transmons, therefore, likewise are expected to support cross-resonance gates and other two-qubit gates applicable to transmons in general.

\noindent \textbf{Acknowledgments} 
B.A., J.B.H. thank Malcolm Carroll, Easwar Magesan, and Jay Gambetta for fruitful discussions. B.A. highly appreciates Jim Rozen's help with wiring the dilution fridge. This work is partially funded by the IARPA Grant No. W911NF-16-1-0114-FE. Contribution of the U.S. Government, not subject to copyright.

\end{document}